\documentclass[a4paper,fleqn,usenatbib]{mnras}

\usepackage{natbib}

\usepackage{savesym}
\usepackage{amsmath}
\savesymbol{iint}
\usepackage{txfonts}
\restoresymbol{TXF}{iint}
\usepackage[T1]{fontenc}
\usepackage{ae,aecompl}

\usepackage{todonotes}
\usepackage{color}

\usepackage{graphicx}	% Including figure files
\usepackage{amssymb}	% Extra maths symbols

\usepackage{natbib}
\newcommand{\lcdm}{\Lambda\mathrm{CDM}}

\newcommand{\fbcen}{f^{\mathrm{blue}}_{\mathrm{cen}}}
\newcommand{\fbsat}{f^{\mathrm{blue}}_{\mathrm{sat}}}

\newcommand{\frcen}{f^{\mathrm{red}}_{\mathrm{cen}}}
\newcommand{\frsat}{f^{\mathrm{red}}_{\mathrm{sat}}}

\newcommand{\tot}{_{\mathrm{tot}}}
\newcommand{\cen}{_{\mathrm{cen}}}

\newcommand{\sat}{_{\mathrm{sat}}}
\newcommand{\red}{^{\mathrm{red}}}
\newcommand{\blue}{^{\mathrm{blue}}}

\newcommand{\avg}[1]{\left\langle #1 \right\rangle}
\newcommand{\dd}{\mathrm{d}}
\newcommand{\hmsol}{h^{-1}M_\odot}
\newcommand{\ms}{M_*}
\newcommand{\msq}{M_*^q}
\newcommand{\mh}{M_h}
\newcommand{\mhq}{M_h^q}
\newcommand{\mhqc}{M_h^{qc}}

\newcommand{\mhqs}{M_h^{qs}}

\newcommand{\muc}{\mu^{c}}
\newcommand{\mus}{\mu^{s}}
\newcommand{\mpc}{\mathrm{Mpc}}
\newcommand{\hmpc}{h^{-1}\mathrm{Mpc}}

\newcommand{\hkpc}{h^{-1}\mathrm{kpc}}

\newcommand{\ihod}{\texttt{iHOD}}
\newcommand{\hhmsol}{h^{-2}M_\odot}
\newcommand{\ds}{\Delta\Sigma}
\newcommand{\kms}{\mathrm{km}\,s^{-1}}
\title[{\ihod} quenching model: separating red and blue galaxies]{Mapping stellar content to dark matter halos. II. Halo mass is the main
driver of galaxy quenching}

\author[Zu \& Mandelbaum 2015]{
Ying  Zu$^{1}$\thanks{E-mail: yzu@cmu.edu}
and Rachel Mandelbaum$^{1}$
\\
$^{1}$McWilliams Center for Cosmology, Department of Physics, Carnegie Mellon University, 5000 Forbes Avenue,
Pittsburgh, PA 15213, USA\\
}

\date{Accepted XXX. Received YYY; in original form ZZZ}

\pubyear{2015}

\begin{document}
\label{firstpage}
\pagerange{\pageref{firstpage}--\pageref{lastpage}}
\maketitle

\begin{abstract}
    We develop a simple yet comprehensive method to distinguish the underlying drivers of galaxy quenching,
    using the clustering and galaxy-galaxy lensing of red and blue galaxies in SDSS. Building on the {\ihod}
    framework developed by \citet{zm15}, we consider two quenching scenarios: 1) a {\it``halo''} quenching
    model in which halo mass is the sole driver for turning off star formation in both centrals and
    satellites; and 2) a {\it ``hybrid''} quenching model in which the quenched fraction of galaxies depends
    on their stellar mass while the satellite quenching has an extra dependence on halo mass. The two best-fit
    models describe the red galaxy clustering and lensing equally well, but halo quenching provides
    significantly better fits to the blue galaxies above $10^{11}\hhmsol$.  The halo quenching model also
    correctly predicts the average halo mass of the red and blue centrals, showing excellent agreement with
    the direct weak lensing measurements of locally brightest galaxies.  Models in which quenching is not tied
    to halo mass, including an age-matching model in which galaxy colour depends on halo age at fixed $M_*$,
    fail to reproduce the observed halo mass for massive blue centrals.  We find similar critical halo masses
    responsible for the quenching of centrals and satellites ~($\sim 1.5\times10^{12}\hmsol$), hinting at a
    uniform quenching mechanism for both, e.g., the virial shock-heating of infalling gas. The success of the
    {\ihod} halo quenching model provides strong evidence that the physical mechanism that quenches star
    formation in galaxies is tied principally to the masses of their dark matter halos rather than the
    properties of their stellar components.
\end{abstract}


%%%%%%%%%%%%%%%%%%%%%%%%%%%%%%%%%%%%%%%%%%%%%%%%%%

%%%%%%%%%%%%%%%%% BODY OF PAPER %%%%%%%%%%%%%%%%%%
\section{Introduction}
\label{sec:intro}

The quenching of galaxies, namely, the relatively abrupt shutdown of star formation activities, gives rise to
two distinctive populations of quiescent and active galaxies, most notably manifested in the strong bimodality
of galaxy colours~\citep{strateva2001, baldry2006}. The underlying driver of quenching, whether it be stellar
mass, halo mass, or environment, should produce an equally distinct split in the spatial clustering and weak
gravitational lensing between the red and blue galaxies.  Recently, \citet[][hereafter Paper I]{zm15}
developed a powerful statistical framework, called the {\ihod} model, to interpret the spatial
clustering~(i.e., the projected galaxy autocorrelation function $w_p$) and the galaxy-galaxy~(g-g)
lensing~(i.e., the projected surface density contrast $\ds$) of the overall galaxy population in the Sloan
Digital Sky Survey~\citep[SDSS;][]{york2000}, while establishing a robust mapping between the observed
distribution of stellar mass to that of the underlying dark matter halos.  In this paper, by introducing two
empirically-motivated and physically-meaningful quenching models within {\ihod}, we hope to robustly identify
the dominant driver of galaxy quenching, while providing a self-consistent framework to explain the bimodality
in the spatial distribution of galaxies.

Galaxies cease to form new stars and become quenched when there is no cold gas. Any physical process
responsible for quenching has to operate in one of three following modes: 1) it heats up the gas to high
temperatures and stops hot gas from cooling efficiently~\citep[e.g., gravitational collapse and various
baryonic feedback; see][for a review]{benson2010}; 2) it depletes the cold gas reservoir via secular stellar
mass growth or sudden removal by external forces~\citep[e.g., tidal and ram pressure;][]{gunn1972}; and 3) it
turns off gas supply by slowly shutting down accretion~\citep[e.g., strangulation;][]{balogh2000}. However,
due to the enormous complexity in the formation history of individual galaxies,
multiple quenching modes may play a role in the history of quiescent galaxies. Therefore, it is more promising
to focus on the underlying physical driver of the {\it average} quenching process, which is eventually tied to
either the dark matter mass of the host halos, the galaxy stellar mass, or the small/large-scale environment
density that the galaxies reside in, hence the so-called ``halo'', ``stellar mass'', and ``environment''
quenching mechanisms, respectively.

Halo quenching has provided one of the most coherent quenching scenarios from the theoretical perspective. In
halos above some critical mass~($M_{\mathrm{shock}}{\sim}10^{12}\hmsol$), virial shocks heat gas inflows from
the intergalactic medium, preventing the accreted gas from directly fueling star formation~\citep{binney1977,
birnboim2003, katz2003, binney2004, keres2005, keres2009}. Additional heating from, e.g., the active galactic
nuclei~(AGNs) then maintains the gas coronae at high temperature~\citep{croton2006}. For halos with
$M_h{<}M_{\mathrm{shock}}$, the incoming gas is never heated to the virial temperature due to rapid post-shock
cooling, therefore penetrating the virial boundary into inner halos as cold flows. This picture, featuring a
sharp switch from the efficient stellar mass buildup via filamentary cold flow into low mass halos, to the
halt of star formation due to quasi-spherical hot-mode accretion in halos above $M_{\mathrm{shock}}$,
naturally explains the colour bimodality, particularly the paucity of galaxies transitioning from blue,
star-forming galaxies to the red sequence of quiescent galaxies~\citep{cattaneo2006, dekel2006}.  To first
order, halo quenching does not discriminate between centrals and satellites, as both are immersed in the same
hot gas coronae that inhibits star formation.  However, since the satellites generally lived in lower mass
halos before their accretion and may have retained some cold gas after accretion, the dependence of satellite
quenching on halo mass should have a softer transition across $M_{\mathrm{shock}}$, unless the quenching by
hot halos is instantaneous.

Observationally, by studying the dependence of the red galaxy fraction $f\red$ on stellar mass $\ms$ and
galaxy environment $\delta_{\mathrm{5NN}}$~(i.e., using distance to the 5th nearest neighbour) in both the
Sloan Digital Sky Survey~(SDSS) and zCOSMOS, \citet[][hereafter P10]{peng2010} found that $f\red$ can be
empirically described by the product of two independent trends with $\ms$ and $\delta_{\mathrm{5NN}}$,
suggesting that stellar mass and environment quenching are at play. By using a group catalogue constructed
from the SDSS spectroscopic sample, \citet{peng2012} further argued that, while the stellar mass quenching is
ubiquitous in both centrals and satellites, environment quenching mainly applies to the satellite galaxies.

However, despite the empirically robust trends revealed in P10, the interpretations for both the stellar mass
and environment trends are obscured by the complex relation between the two observables and other physical
quantities.  In particular, since the observed $\ms$ of central galaxies is tightly correlated with halo mass
$\mh$~(with a scatter ${\sim}0.22$ dex; see Paper I), a stellar mass trend of $f\red$ is almost
indistinguishable with an underlying trend with halo mass.  By examining the inter-relation among $\ms$,
$\mh$, and $\delta_{\mathrm{5NN}}$, \citet{woo2013} found that the quenched fraction is more strongly
correlated with $\mh$ at fixed $\ms$ than with $\ms$ at $\mh$, and the satellite quenching by
$\delta_{\mathrm{5NN}}$ can be re-interpreted as halo quenching by taking into account the dependence of
quenched fraction on the distances to the halo centres. The halo quenching interpretation of the stellar and
environment quenching trends is further demonstrated by \citet{gabor2015}, who implemented halo quenching in
cosmological hydrodynamic simulations by triggering quenching in regions dominated by hot~($10^{5.4} K$) gas.
They reproduced a broad range of empirical trends detected in P10 and \citet{woo2013}, suggesting that the
halo mass remains the determining factor in the quenching of low-redshift galaxies.

Another alternative quenching model is the so-called ``age-matching'' prescription of \citet{hearin2013} and
its recently updated version of \citet{hearin2014}.  Age-matching is an extension of the ``subhalo abundance
matching''~\citep[SHAM;][]{conroy2006} technique, which assigns stellar masses to individual
subhalos~(including both main and subhalos) in the N-body simulations based on halo properties like the peak
circular velocity~\citep{reddick2013}.  In practice, after assigning $\ms$ using SHAM, the age-matching method
further matches the colours of galaxies at fixed $\ms$ to the ages of their matched halos, so that older halos
host redder galaxies.  In essence, the age-matching prescription effectively assumes a stellar mass quenching,
as the colour assignment is done at fixed $\ms$ regardless of halo mass or environment, with a secondary
quenching via halo formation time.  Therefore, the age-matching quenching is very similar to the
$\ms$-dominated quenching of P10, except that the second variable is halo formation time rather than galaxy
environment.

The key difference between the $\mh$- and $\ms$-dominated quenching scenarios lies in the way central galaxies
become quiescent. One relies on the stellar mass while the other on the mass of the host halos, producing two
very different sets of colour-segregated stellar-to-halo relations~(SHMRs). At fixed halo mass, if
stellar mass quenching dominates, the red centrals should have a higher average stellar mass than the blue
centrals; in the halo quenching scenario the two coloured populations at fixed halo mass would have
similar average stellar masses, but there is still a trend for massive galaxies to be red because higher mass
halos host more massive galaxies. This difference in SHMRs directly translates to two distinctive ways the
red and blue galaxies populate the underlying dark matter halos according to their $\ms$ and $\mh$, hence two
different spatial distributions of galaxy colours.

Therefore, by comparing the $w_p$ and $\ds$ predicted from each quenching model to the measurements from SDSS,
we expect to robustly distinguish the two quenching scenarios.  The {\ihod} framework we developed in Paper I
is ideally suited for this task.  The {\ihod} is a global ``halo occupation distribution''~(HOD) model defined on
a 2D grid of $\ms$ and $\mh$, which is crucial to modelling the segregation of red and blue galaxies in their
$\ms$ distributions at fixed $\mh$.  The {\ihod} quenching constraint is fundamentally different and
ultimately more meaningful compared to approaches in which colour-segregated populations are treated
independently~\citep[e.g.,][]{tinker2013, puebla2015}. Our {\ihod} quenching model automatically fulfills the
consistency relation which requires that the sum of red and blue SHMRs is mathematically identical to the
overall SHMR. More importantly, the {\ihod} quenching model employs only four additional parameters that are
directly related to the average galaxy quenching, while most of the traditional approaches require ${\sim}20$
additional parameters, rendering the interpretation of constraints difficult. Furthermore, the
{\ihod} framework allows us to include ${\sim}80\%$ more galaxies than the traditional HODs and take into
account the incompleteness of stellar mass samples in a self-consistent manner.

This paper is organized as follows. We describe the selection of red and blue samples in
Section~\ref{sec:data}. In Section~\ref{sec:model} we introduce the parameterisations of the two quenching
models and derive the {\ihod}s for each colour. We also briefly describe the signal measurement and model
prediction in Sections~\ref{sec:data} and ~\ref{sec:model}, respectively, but refer readers to Paper I for
more details. The constraints from both quenching mode analyses are presented in
Section~\ref{sec:constraint}. We perform a thorough model comparison using two independent criteria in
Section~\ref{sec:result} and discover that halo quenching model is strongly favored by the data. In
Section~\ref{sec:physics} we discuss the physical implications of the halo quenching model and compare it to
other works in~\ref{sec:compare}. We conclude by summarising our key findings in Section~\ref{sec:conclusion}.

Throughout this paper and Paper I, we assume a $\lcdm$ cosmology with $(\Omega_m, \Omega_{\Lambda}, \sigma_8,
h)\,{=}\,(0.26, 0.74, 0.77, 0.72)$.  All the length and mass units in this paper are scaled as if the Hubble
constant were $100\,\kms\mpc^{-1}$. In particular, all the separations are co-moving distances in units of
either $\hkpc$ or $\hmpc$, and the stellar mass and halo mass are in units of $\hhmsol$ and $\hmsol$,
respectively.  Unless otherwise noted, the halo mass is defined by
$\mh\,{\equiv}\,M_{200m}\,{=}\,200\bar{\rho}_m(4\pi/3)r_{200m}^3$, where $r_{200m}$ is the corresponding halo
radius within which the average density of the enclosed mass is $200$ times the mean matter density of the
Universe, $\bar{\rho}_m$.  For the sake of simplicity, $\ln x{=}\log_e x$ is used for the natural logarithm,
and $\lg x{=}\log_{10} x$ is used for the base-$10$ logarithm.

\section{Sample Selection and Signal Measurement}
\label{sec:data}

In this section we describe the SDSS data used in this paper, especially the selection of the red and blue
galaxies within the stellar mass samples, and the measurements of the galaxy clustering and the g-g lensing
signals. We briefly describe the overall large-scale structure sample and the signal measurement, same as that
used in Paper I, below in Section~\ref{subsec:vagc} and~\ref{subsec:signals}, respectively, and refer readers
to Paper I for details.  Here we focus more on the colour cut we employ to divide the galaxies into red and
blue populations in Section~\ref{subsec:colourcut}.

\subsection{NYU--VAGC and Stellar Mass Samples}
\label{subsec:vagc}

\begin{figure*}
\begin{center}
\includegraphics[width=1\textwidth]{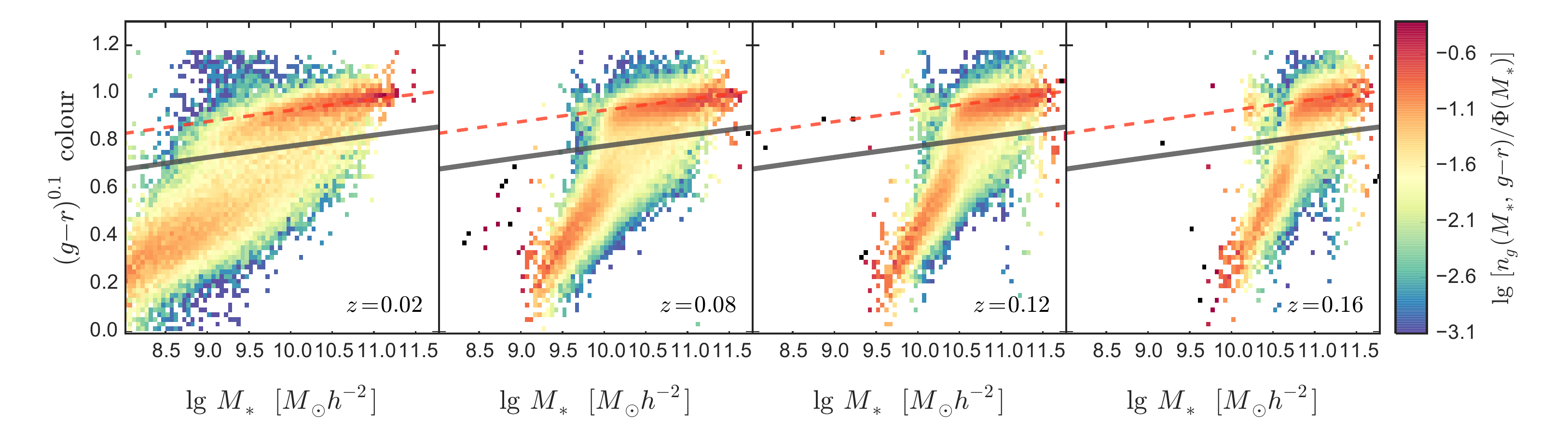}
\caption[]{\input{figures/cmd/caption.tex}
}
\end{center}
\end{figure*}

We make use of the final data release of the SDSS~\citep[DR7;][]{abazajian2009}, which contains the completed
data set of the SDSS-I and the SDSS-II. In particular, we obtain the Main Galaxy Sample~(MGS) data from the
\texttt{dr72} large--scale structure sample \texttt{bright0} of the ``New York University Value Added
Catalogue''~(NYU--VAGC), constructed as described in~\citet{blanton2005}. The \texttt{bright0} sample includes
galaxies with $10{<}m_r{<}17.6$, where $m_r$ is the $r$-band Petrosian apparent magnitude, corrected for
Galactic extinction. We apply the ``nearest-neighbour'' scheme to correct for the $7\%$ galaxies that are
without redshift due to fibre collision, and use data exclusively within the contiguous area in the North
Galactic Cap and regions with angular completeness greater than $0.8$. The final sample used for the galaxy
clustering analysis includes $513{,}150$ galaxies over a sky area of $6395.49$ deg$^2$. A further 5 per cent
of the area is eliminated for the lensing analysis, due to the absence of source galaxies in that area.

As discussed in Paper I, we further restrict our analysis to galaxies above a ``mixture limit'', defined as
the stellar mass threshold above which the galaxy sample is relatively complete with a fair mix of red
and blue galaxies.  The functional form we adopt to describe the mixture limit $M_*^{\mathrm{mix}}(z)$ is
\begin{equation}
    \lg\left(\frac{M_*^{\mathrm{mix}}}{\hhmsol}\right) = 5.4 \times (z-0.025)^{0.33} + 8.0, \label{eqn:ql}
\end{equation}
shown as the thick yellow curves in Fig.~\ref{fig:samsel}~(discussed further below).
By taking into account the sample incompleteness in a self-consistent way,
the {\ihod} model is able to model the lensing and clustering statistics of all galaxies above the
mixture limit, ${\sim}84\%$ more than the traditional HOD
models typically include from the same catalogue.

We employ the stellar mass estimates from the latest MPA/JHU value-added galaxy
catalogue\footnote{\url{http://home.strw.leidenuniv.nl/~jarle/SDSS/}}. The stellar masses were estimated based
on fits to the SDSS photometry following the philosophy of~\citet{kauffmann2003} and~\citet{salim2007}, and
assuming the Chabrier~\citep{chabrier2003} initial mass function~(IMF) and the~\citet{bruzual2003} SPS model.
The MPA/JHU stellar mass catalogue is then matched to the NYU-VAGC \texttt{bright0} sample. We identify
valid, unambiguous MPA/JHU stellar mass estimates for all but $32{,}327$~($6.3\%$) of the MGS galaxies. For those
unmatched galaxies, we predict their stellar masses using the overall scaling between the two stellar mass
estimates, depending on the $g{-}r$ colours~(k-corrected to $z{=}0.1$).

As sources for the g-g lensing measurement, we use a catalogue of background galaxies
\citep{2012MNRAS.425.2610R} with a number density of 1.2 arcmin$^{-2}$ with weak lensing shears estimated
using the re-Gaussianization method \citep{2003MNRAS.343..459H} and photometric redshifts from Zurich
Extragalactic Bayesian Redshift Analyzer \citep[ZEBRA,][]{2006MNRAS.372..565F}.   The catalogue was
characterised in several papers that describe the data, and use both the data and simulations to estimate
systematic errors \citep[see ][]{2012MNRAS.425.2610R,2012MNRAS.420.1518M,2012MNRAS.420.3240N,mandelbaum2013}.

\subsection{Separating Sample into Red and Blue}
\label{subsec:colourcut}

\begin{figure*}
\begin{center}
\includegraphics[width=1\textwidth]{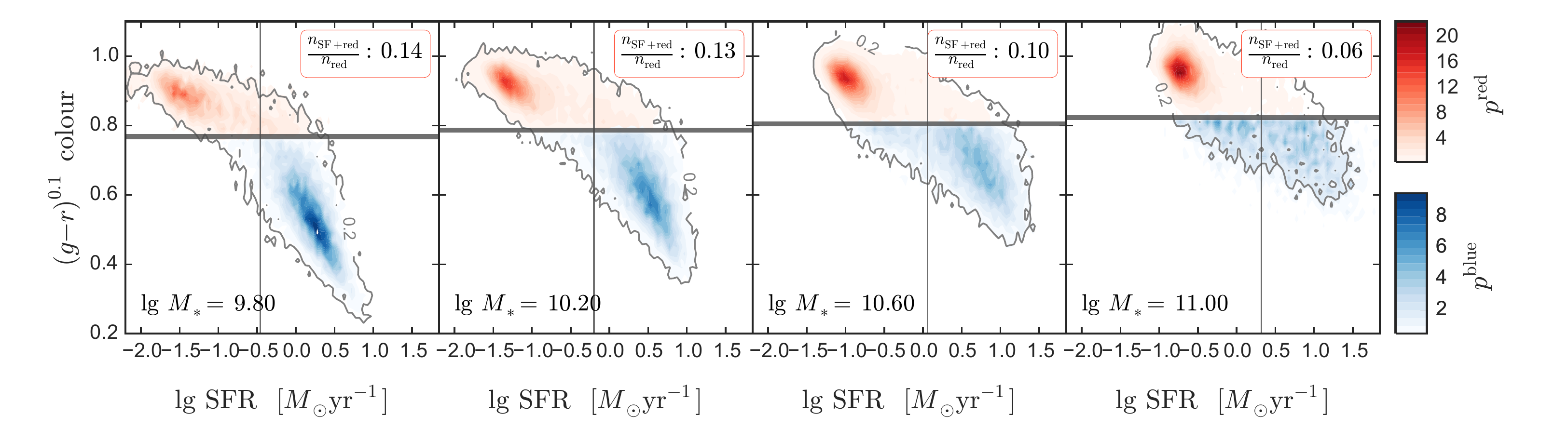}
\caption[]{\input{figures/sfr/caption.tex}
}
\end{center}
\end{figure*}

We define quenching by the $(g{-}r)^{0.1}$ colour~(after K-correction to $z{=}0.1$) for three reasons: 1)
colour bimodality is very stable across different environments and redshifts~\citep{baldry2006}; 2)
observationally colour is very easy to measure robustly, without the need to fit galaxy morphology or
brightness profile; and 3) physically colour is the result of integrated star formation history, largely
immune to incidental or minor star formation episodes. In addition, we aim to model and compare to the two
separable quenching trends with stellar mass and environment that revealed in P10, who originally chose
optical colour as the quenching indicator.

\begin{figure*}
\begin{center}
\includegraphics[width=0.8\textwidth]{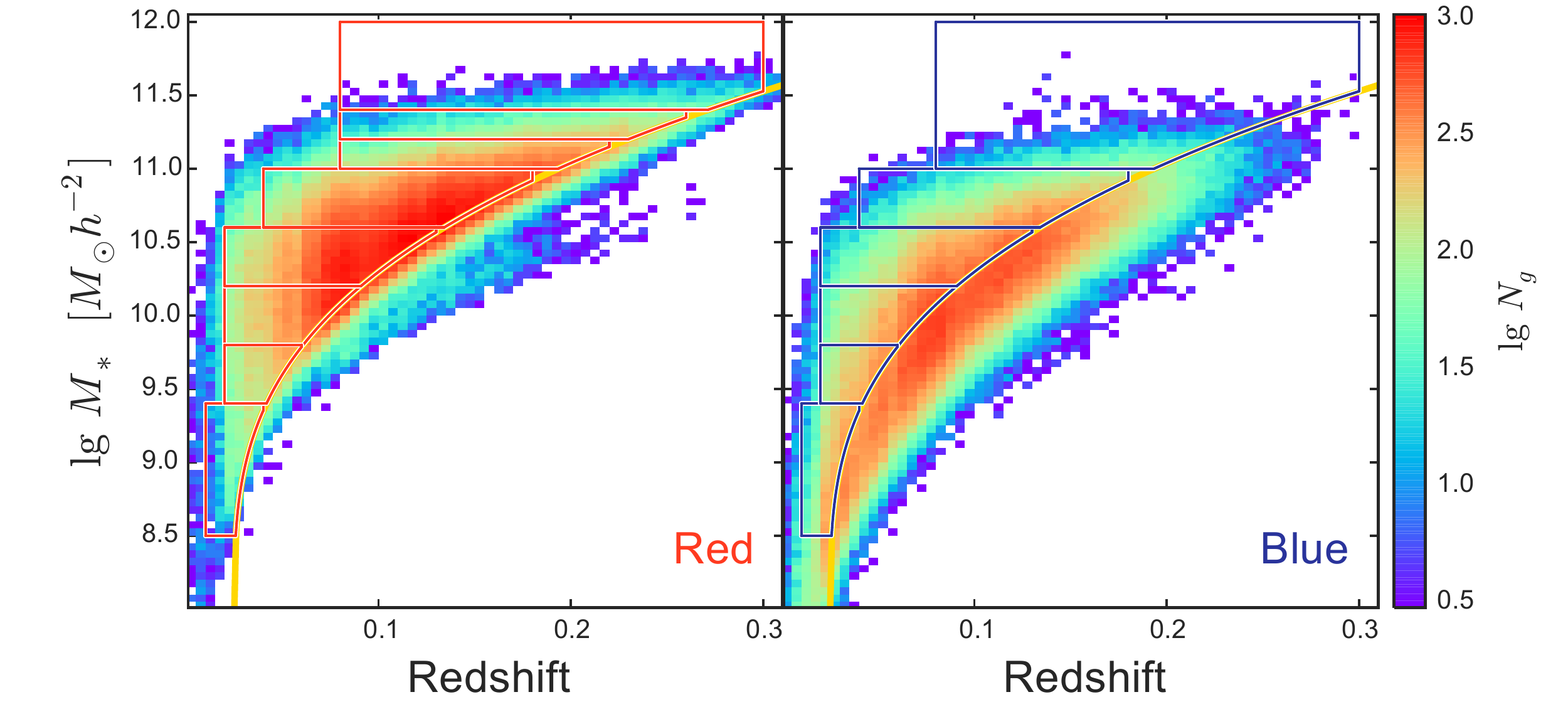}
\caption[]{\input{figures/samsel/caption.tex}
}
\end{center}
\end{figure*}

Fig.~\ref{fig:cmd} illustrates the colour-stellar mass diagrams~(CSMDs) at four different redshifts and the
stellar mass dependence of the colour cuts we applied to divide the red and blue galaxies. In each panel, the colour map
indicates the distribution of the logarithmic comoving number density of galaxies in cells of $(g{-}r)^{0.1}$
and $\ms$, normalized by the stellar mass function~(SMF) at that $\ms$. This normalization highlights the
$g{-}r$ ranges with relatively high concentration of galaxies along the colour axis, enhancing the appearance
of the ``red sequence''  on each panel. The CSMDs are cut off at different stellar masses due to the
redshift dependence of the mixture limit, which is ultimately related to the flux limit of the spectroscopic
survey.  The red dashed lines going through the red sequences are uniform across all redshifts, indicating
that the loci of the red sequence on the CSMD is independent of redshift within our sample.
To divide the galaxies into red and blue, we therefore define
the colour cut to be parallel to the red sequence on the CSMD, described by
\begin{equation}
\left(g-r\right)^{0.1}_{\mathrm{cut}}(\ms) = 0.8 \left(\frac{\lg\ms}{10.5}\right)^{0.6},
\label{eqn:cut}
\end{equation}
and indicated by the black solid lines in Fig.~\ref{fig:cmd}. The weak stellar mass dependence in
equation~\ref{eqn:cut} causes a variation of $(g-r)^{0.1}_{\mathrm{cut}}$ between $0.76$ and $0.84$ within our
sample, leading to differences in classification between blue vs. red of only a few per cent compared to a
constant cut at $0.8$.

Whether a galaxy is quiescent or star-forming, however, is never a clear-cut choice. Galaxy bimodality shows
in nearly every aspect of galaxy properties, including broad-band colour, star formation rate~(SFR),
morphology~(e.g., late/early-type,  De Vaucouleurs/exponential profile), and concentration~(e.g., S\'ersic
index). \citet{bernardi2010} found that many late-type (Sb and later) galaxies lie above the red galaxy colour
cut~(similar to equation~\ref{eqn:cut}) and they tend to be edge-on discs reddened by dust. Conversely, some
early-type galaxies lie below the cut, either showing star-forming AGN or post-starburst spectrum, with their
star formation history well described by a recent minor and short starburst superimposed on old stellar
component~\citep{huang2009}. As a result, \citet{woo2013} advocated the use of SFR as the quenching indicator,
and they claimed that one third of the red galaxies are star forming. However, the large fraction of
star-forming contaminants in the ``red'' population in~\citet{woo2013} is mainly caused by the rest-frame
$U{-}B$ colour \citet{woo2013} adopted in defining red galaxies, derived from $AB$ magnitudes that are
K-corrected from the SDSS $ugriz$ photometry. Using the native $(g{-}r)^{0.1}$ colour largely eliminates the
star-formers from the red galaxies.

Fig.~\ref{fig:sfr} clearly demonstrates the good consistency between using $(g{-}r)^{0.1}$ colour and SFR as
quenching indicators. The four panels show the joint 2D probability density distributions~(PDFs) of galaxy
colour and the logarithmic SFR at four different $\ms$~(from left to right: $\lg\ms=9.8$, $10.2$, $10.6$, and
$11.0$). In each panel, the thick horizontal line represents the colour cut defined in equation~\ref{eqn:cut},
while the thin vertical line indicates the SFR value that saddles the separate SFR distributions of passive and
active galaxies, which can be well described by
\begin{equation}
\lg\,\frac{\mathrm{SFR}_{\mathrm{cut}}}{\ms} = -0.35 (\lg\ms-10.0)- 10.23,
\label{eqn:sfrcut}
\end{equation}
in parallel to the star-forming sequence defined in \citet{salim2007}. The fraction of dusty,
star-forming galaxies in the red population decreases from $0.14$ to $0.06$ as stellar mass increases from
$\lg\ms{=}9.8$ to $11.0$, significantly lower than the one third reported in \citet{woo2013} using $U{-}B$
colours. Therefore, we conclude that it is robust to use red fraction as a proxy for quenching efficiency,
and the results of our analysis should stay the same if SFR were used for the selection of quenched galaxies.

As described in Paper I, the {\ihod} model constructs individual HODs within very narrow redshift slices~(we
use $\Delta z{=}0.01$), so that the sample selection does not require a single uniform stellar mass range
among all the redshift slices within that sample, i.e., having a rectangular shape on the $\ms$-$z$ diagrams.
Fig.~\ref{fig:samsel} illustrates the galaxy samples selected on the $\ms$-$z$ diagram within each coloured
population for the {\ihod} quenching analysis. The colour intensity represents the logarithmic galaxy number
counts in cells of $\ms$ and $z$. As mentioned in Section~\ref{subsec:vagc}, all selected samples have the
``wedge''-like stellar mass thresholds described by the mixture limit, and thus contain extra galaxies at the
far end of the redshift range that are usually unused in traditional HOD analysis. Additionally, since those
high redshift wedges have a larger comoving volume per unit redshift than the low redshifts, they include the
most abundant regions on the $\ms$-$z$ diagram, corresponding to the reddest regions on both panels of
Fig.~\ref{fig:samsel}. The resultant increase in the selected galaxy sample sizes is more than $80\%$ compared
to traditional selections.

Above the mixture limit, the red galaxies~(left panel) are two times more abundant than the blue
galaxies~(right panel), despite the fact that the ratio of the two colours is close to unity in the
spectroscopic survey.  Therefore, we can afford finer binning in $\ms$ in the red galaxy sample than in the
blue galaxy sample, especially at the high mass end.  Table~\ref{tab:smbins} summarises the basic information
of the two sets of sample selections used by the {\ihod} quenching analysis.  In total, we divide $228{,}759$
red galaxies and $94{,}325$ blue galaxies into eight and six subsamples, respectively.

\begin{table}
\centering \caption{Red and blue stellar mass bins used for the {\ihod} quenching analysis, corresponding
    to the selections in the left and right panels of Figure~\ref{fig:samsel}, respectively.  The
    red selection includes three stellar mass samples above $\lg\ms{=}11$ as opposed to the single
    $\lg\ms{>}11$ sample in the blue selection, while the five lower stellar
    mass red and blue samples share the same binning in $\lg\ms$ and $z$.
}
\begin{tabular}{cccc}
\hline
\hline
$\lg(\ms/\hhmsol)$  & $z$   & $N\red$  & $N\blue$ \\
\hline
 8.5$-$9.4  & 0.01$-$0.04 & $  3{,}224$  & $10{,}773 $ \\
 9.4$-$9.8  & 0.02$-$0.06 & $  7{,}336$  & $ 9{,}356 $ \\
 9.8$-$10.2 & 0.02$-$0.09 & $ 28{,}301$  & $19{,}883 $ \\
10.2$-$10.6 & 0.02$-$0.13 & $ 70{,}514$  & $29{,}160 $ \\
10.6$-$11.0 & 0.04$-$0.18 & $ 84{,}108$  & $21{,}058 $ \\
11.0$-$11.2 & 0.08$-$0.22 & $ 22{,}626$  &             \\
11.2$-$11.4 & 0.08$-$0.26 & $  9{,}775$  &             \\
11.4$-$12.0 & 0.08$-$0.30 & $  2{,}875$  &             \\
11.0$-$12.0 & 0.08$-$0.30 &              & $4{,}095$ \\
\hline
\end{tabular}
\label{tab:smbins}
\end{table}

\subsection{Measuring Galaxy Clustering $w_p$ and Galaxy-Galaxy Lensing $\ds$ }
\label{subsec:signals}

We measure the projected correlation function $w_p$ for each galaxy sample by integrating the 2D
redshift--space correlation function $\xi^s$,
\begin{equation}
    w_p(r_p) = \int_{-r_{\pi}^{\mathrm{max}}}^{+r_{\pi}^{\mathrm{max}}} \xi^{s}(r_p, r_\pi) d r_\pi,
    \label{eqn:wp}
\end{equation}
where $r_p$ and $r_\pi$ are the projected and the line-of-sight~(LOS) comoving distances between two
galaxies. We measure the $w_p$ signal out to a maximum projected distance of $r_p^{\mathrm{max}}=20\hmpc$,
where the galaxy bias is approximately linear. For the integration limit, we adopt a maximum LOS distance of
$r_{\pi}^{\mathrm{max}}{=}60\hmpc$. We only use the $w_p$ values down to the physical distance that
corresponds to the fibre radius at the maximum redshift of each sample, with fewer data points for higher
stellar mass~(hence larger maximum fibre radius) samples.

The Landy--Szalay estimator~\citep{landy1993} is employed for computing the 2D correlation $\xi^{s}(r_p, r_\pi)$
The error covariance matrix for each $w_p$ measurement is estimated via the jackknife re-sampling technique. We
divide the entire footprint into $200$ spatially contiguous, roughly equal--size patches on the sky and
compute the $w_p$ for each of the $200$ jackknife subsamples by leaving out one patch at a time. For each
stellar mass sample, we adopt the sample mean of the $200$ subsample measurements as our final estimate of
$w_p$, and the sample covariance matrix as an approximate to the underlying error covariance.

For the surface density contrast $\ds$,  we measure the projected mass density in each radial bin by summing
over lens-source pairs ``$ls$'' and random lens-source pairs ``$rs$'',
\begin{equation}\label{eq:ds-estimator}
\Delta\Sigma(r_p) = \langle \Sigma_{{\rm crit}}\gamma_t(r_p) \rangle = \frac{\sum_{ls} w_{ls} e_t^{(ls)} \Sigma_{{\rm crit}}(z_l,z_s)}{2 {\cal R}\sum_{rs} w_{rs}},
\end{equation}
where $e_t$ is the tangential ellipticity component of the source galaxy with respect to the lens position,
the factor of $2{\cal R}$ converts our definition of ellipticity to the tangential shear $\gamma_t$,
and $w_{ls}$ is the inverse variance weight assigned to each lens-source pair~(including shot noise and
measurement error terms in the variance). $\Sigma_{\rm crit}$ is the so-called critical surface mass density,
defined as
\begin{equation}
\Sigma_{\rm
crit}^{-1}(z_l,z_s)\equiv \frac{4\pi G}{ c^{2}}\frac{D_{ls}D_l(1+z_l)^2}{D_s},
\end{equation}
where $D_l$ and $D_s$ are the angular diameter distances to lens and source, and $D_{ls}$ is the distance
between them.  We use the estimated photometric redshift each source to compute $D_s$ and $D_{ls}$.  The
factor of $(1+z_l)^2$ comes from our use of comoving coordinates. We subtract off a similar signal measured
around random lenses, to subtract off any coherent systematic shear contributions \citep{2005MNRAS.361.1287M};
this signal is statistically consistent with zero for all scales used in this work. Finally, we correct a bias
in the signal caused by the uncertainties in the photometric redshift using the method from
\cite{2012MNRAS.420.3240N}.

To calculate the error bars, we also used the jackknife re-sampling method.  As shown in
\cite{2005MNRAS.361.1287M}, internal estimators of error bars (in that case, bootstrap rather than jackknife)
perform consistently with external estimators of errorbars for $\Delta\Sigma$ on small scales due to its being
dominated by shape noise.

\section{Quenching Models and Signal Predictions}
\label{sec:model}

\begin{figure*}
\begin{center}
\includegraphics[width=0.8\textwidth]{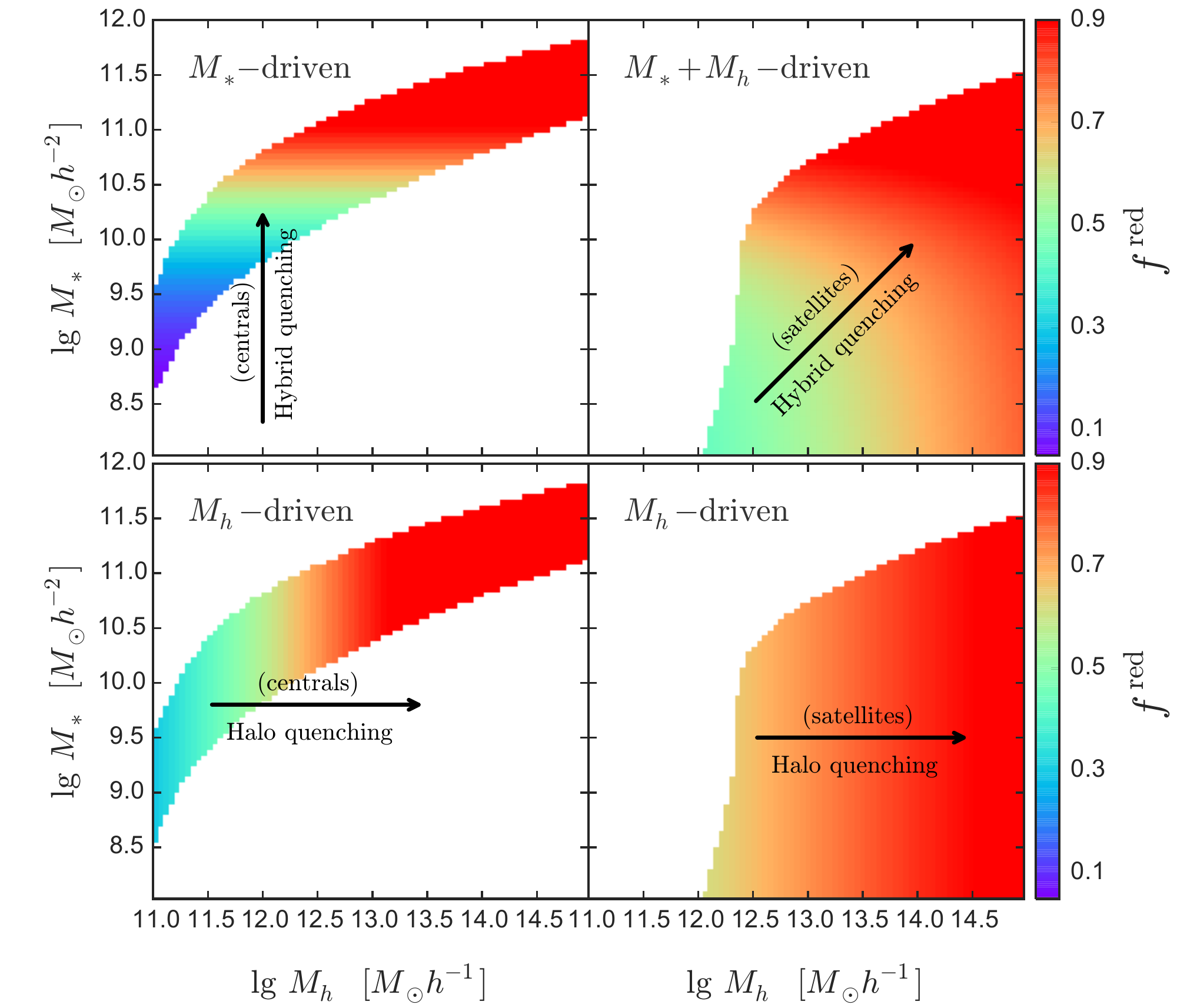}
\caption[]{\input{figures/qmodels/caption.tex}
}
\end{center}
\end{figure*}

In this section, we introduce the mathematical descriptions of the hybrid~(Section~\ref{subsec:hybrid}) and
the halo~(Section~\ref{subsec:halo}) quenching models. We also briefly describe how to infer the {\ihod}s for
the red and blue galaxies in Section~\ref{subsec:ihod}, but refer reader to Paper I for more details on the
{\ihod} framework. The prediction of $w_p$ and $\ds$ from each coloured {\ihod} is rather complex but exactly
the same as that for the overall galaxy populations, therefore we directly refer readers to the relevant
sections~(4 and 5) in Paper I for details. We ignore quenching via mergers in both quenching models considered
below, as merging-induced quenching is negligible at $z{<}0.5$~\citep{peng2010}.

\subsection{Hybrid Quenching Model}
\label{subsec:hybrid}

The hybrid quenching model parameterizes the red fraction as a function of both $\ms$ and $\mh$, aiming to
mimic the empirical stellar mass and environment quenching trends observed in P10. In the physical picture
implied by this model, every quiescent galaxy had spent some portion of its life on the star-forming ``main
sequencing'' as a central before the eventual quenching~\citep{daddi2007, noeske2007, speagle2014}, due to
either the depletion of gas supply~(i.e., stellar mass quenching) or entering another halo as a satellite,
when environment quenching kicked in.

While the stellar mass trend is straightforward to parameterize, it is unclear whether the environment
quenching trend among satellites can be mimicked by a trend in halo mass, as the environment--halo mass
relation is very complex and depends strongly on the definition of that environment.  The P10 environment of
satellites, as defined by $\sigma_{\mathrm{5NN}}$, shows strong correlation with group richness when the
richness is below five. In richer systems, however, the correlation is mostly smeared out and
$\sigma_{\mathrm{5NN}}$ instead anti-correlates with the halo-centric distance $D_g$~\citep{peng2012}.  This
apparent transition between the two richness regimes is caused by the increase of
$D_{\mathrm{5}}/R_{\mathrm{vir}}$, the ratio between the typical distance to the fifth nearest neighbour to
the halo virial radius, from below to above unity. When $D_{\mathrm{5}}>R_{\mathrm{vir}}$,
$\sigma_{\mathrm{5NN}}$ is roughly proportional to $\mh/D_{\mathrm{5}}^3$, thus more tied to $\mh$. When
$D_{\mathrm{5}}<R_{\mathrm{vir}}$, $\sigma_{\mathrm{5NN}}$ is essentially an intra-halo overdensity measured
at $D_g$, which depends more strongly on $D_g$ than $\mh$ due to the steep declining slope of the NFW-like
halo density profile.

But for our purposes, what matters is the {\it mean} satellite quenching efficiency as a function of halo mass
$\mh$, averaged over galaxies at all $D_g$ within that halo.  As pointed out by \citet{woo2013}, the density
profile of more massive halos falls off less steeply with distance than that of less massive systems, so the
probability of finding the $5$-th nearest neighbour increases with halo mass. Therefore, the P10 environment
quenching trend can be potentially encapsulated within the halo model as a satellite quenching dependence on
halo mass.

Assuming stellar mass as the main driver of central galaxy quenching, we parameterize the red fraction of
centrals as
\begin{equation}
    \frcen(\ms, \mh) \equiv 1 - g(\ms) = 1 - \exp\left[-\left( \ms/\msq\right)^{\mu}\right],
    \label{eqn:frcenhybrid}
\end{equation}
where $\msq$ is a characteristic stellar mass~($\frcen(\msq, \mh)=(e-1)/e=0.632$) and $\mu$ dictates how
fast the quenching efficiency increases with $\ms$, with $\mu=1$ being exponential.  The satellites are
subject to an extra halo quenching term $h(\mh)$, so that
\begin{equation}
    \frsat(\ms, \mh) = 1 - g(\ms) h(\mh),
    \label{eqn:frsathybrid}
\end{equation}
with
\begin{equation}
    h(\mh) = \exp\left[-\left( \mh/\mhq\right)^{\nu}\right],
\end{equation}
where $\mhq$ is a characteristic halo mass and $\nu$ controls the pace of satellite quenching. The above
equations, including $g$ and $h$ as powered exponential functions, are very similar to the fitting formula
adopted in \citet{baldry2006} and \citet{peng2012}.

The top left and right panels of Fig.~\ref{fig:qmodels} illustrate the central and satellite red fractions,
computed from the best-fit hybrid quenching model via Equations~\ref{eqn:frcenhybrid}
and~\ref{eqn:frsathybrid}, respectively. The arrow in each panel points in the direction of increasing
quenching efficiency on the 2D plane of $\ms$ and $\mh$. For the central galaxies, although the quenching is
driven by $\ms$ along the horizontal axis, the red fraction still shows strong increasing trend with $\mh$ due
to the tight correlation between $\ms$ and $\mh$, i.e., the SHMR of central galaxies. The 2D distribution of
satellite red fractions displays a ``boxy'' pattern, echoing the separate stellar mass and environment
quenching trends detected in P10.

Combining the central and the satellite terms, the
red fraction of galaxies with stellar mass $\ms$ inside halos of total mass $\mh$ is
\begin{align}
    f\red(\ms, \mh) &= f\sat(\ms, \mh) \; \frsat(\ms, \mh)\;+\nonumber\\
    [1 &- f\sat(\ms, \mh)]\; \frcen(\ms, \mh),
    \label{eqn:fred2d}
\end{align}
where $f\sat(\ms, \mh)$ is the satellite fraction that can be predicted by the overall {\ihod} model from
Paper I. For the hybrid model, equation~\ref{eqn:fred2d} can be reduced to
\begin{equation}
    f\red(\ms, \mh)  = g(\ms)\{1 - f\sat(\ms, \mh)[1-h(\mh)]\}.
\end{equation}

\subsection{The Halo Quenching Model}
\label{subsec:halo}

As described in the introduction, the halo quenching model relies on halo mass alone to quench both central
and satellite galaxies, and \citet{gabor2015} demonstrated that it also naturally explains the stellar mass
and environment quenching trends seen in P10, by embedding galaxies in massive halos filled with hot gas via
virial heating.  However, depending on the exact physical processes driven by $\mh$, halo quenching may apply
to the central and satellites differently. For instance, while the central galaxies in halos above
$M_{\mathrm{shock}}$ could be quenched by shocked-heated gas and then maintain a high gas temperature via the
``raido''-mode feedback from AGNs, the satellite galaxies in the those halos may still retain some cold gas as
the ``central'' galaxies of its own coherent sub-group. Therefore, the satellite galaxies continue to accrete
gas and convert it to stars over a period of ${\sim}1$ Gyr after entering into a larger
halo~\citep{simha2009}. Similar processes like slow strangulation~(assuming no accretion onto satellites) also
produce prolonged quenching actions~\citep{peng2015}.  In this case, the halo quenching of centrals and
satellites are somewhat decoupled, and the quenching of satellites is a more gradual process than that of
centrals.

Therefore, unlike the hybrid model, we describe the red fractions of centrals and satellites as two
independent functions of $\mh$:
\begin{align}
    \frcen(\ms, \mh) &= 1 - \fbcen(\mh) \nonumber\\
    &= 1 - \exp\left[-\left( \mh/\mhqc\right)^{\muc}\right],
    \label{eqn:frcenhalo}
\end{align}
and
\begin{align}
    \frsat(\ms, \mh) &= 1 - \fbsat(\mh) \nonumber\\
    &= 1 - \exp\left[-\left( \mh/\mhqs\right)^{\mus}\right],
    \label{eqn:frsathalo}
\end{align}
where $\mhqc$ and $\mhqs$ are the critical halo masses responsible for triggering quenching of central and
satellites, respectively, and $\muc$ and $\mus$ are the respective powered-exponential indices controlling the
transitional behavior of halo quenching across the critical halo masses.

Similarly, the bottom left and right panels of Fig.~\ref{fig:qmodels} illustrate the central and satellite red
fractions, computed from the best-fit halo quenching model via Equations~\ref{eqn:frcenhalo}
and~\ref{eqn:frsathalo}, respectively, with arrows indicating halo mass as the sole driver for quenching in
both populations. The orthogonality of the hybrid and halo quenching directions for central galaxies is the
key distinction that we look to exploit in this paper, by identifying its imprint on the clustering and g-g
lensing signals of red and blue galaxies. The total red fraction can be obtained by substituting
equations~\ref{eqn:frcenhalo} and ~\ref{eqn:frsathalo} into equation~\ref{eqn:fred2d}.

Finally, we emphasize that in reality the true quenching arrow could be pointing anywhere between the two
orthogonal directions, i.e., a more generalized quenching model consisting of a linear mixture of the two, with
the linear coefficients varying as functions of $\ms$ and $\mh$ as well --- schematically,
\begin{equation}
    Q_{\mathrm{true}} = \omega(\ms, \mh) \times Q_{\mathrm{hybrid}} + (1 - \omega(\ms, \mh)) \times Q_{\mathrm{halo}}.
\end{equation}
However, as a first step of constraining quenching, the goal of this paper is to find out if $\omega$ is
closer to zero~(i.e., halo-quenching dominated) or unity~(i.e, hybrid-quenching dominated).

\begin{figure*}
\begin{center}
\includegraphics[width=0.8\textwidth]{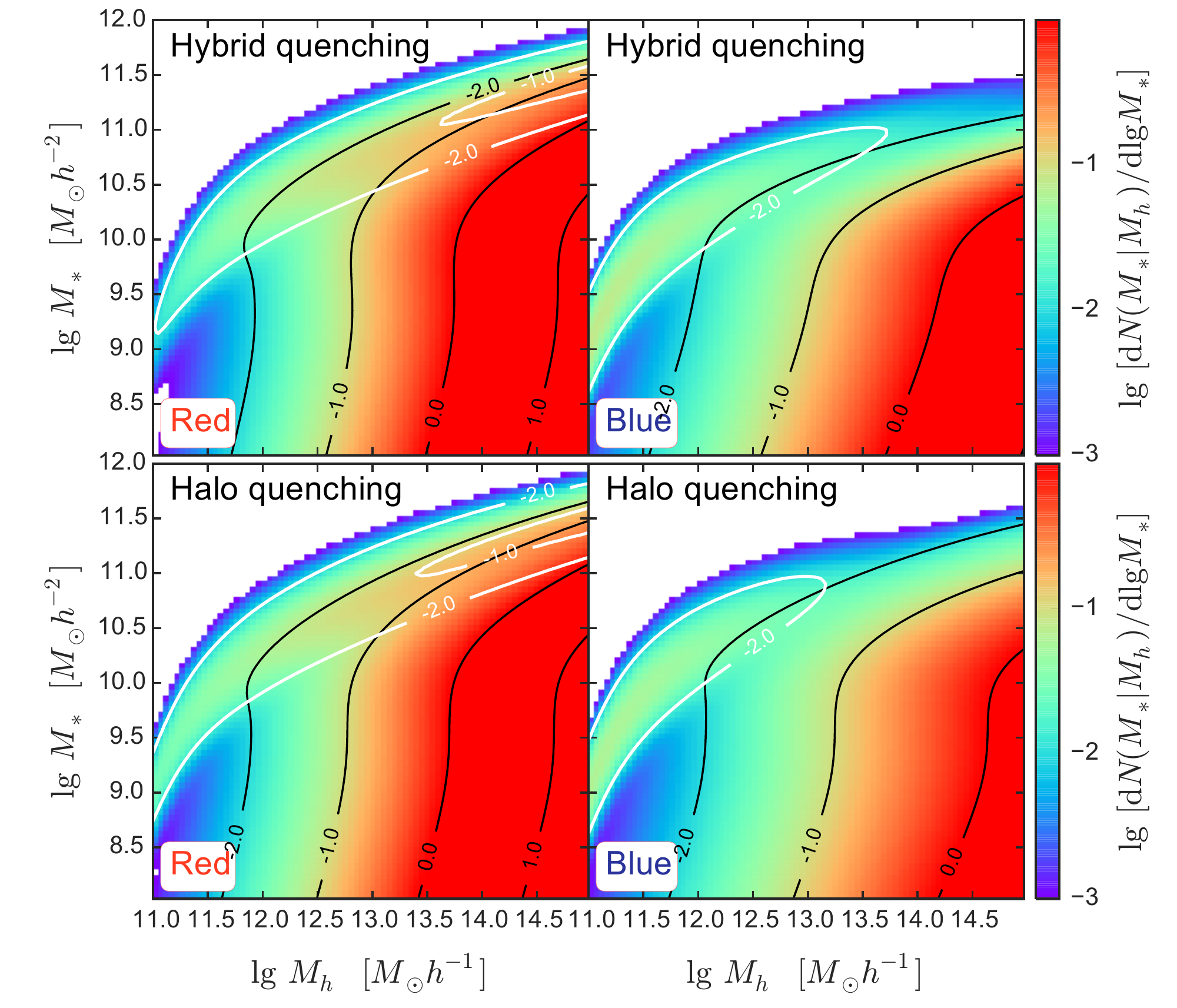}
\caption[]{\input{figures/qihods/caption.tex}
}
\end{center}
\end{figure*}

\subsection{From Quenching Models to Colour-segregated {\ihod}s}
\label{subsec:ihod}

In order to predict the $w_p$ and $\ds$ for the red and blue galaxies in each quenching model, we construct
{\ihod}s for both coloured populations by combining the overall {\ihod} with $f\red(\ms, \mh)$ predicted by
that quenching model.

Let us start with the red galaxies. The key is to derive $p\red(\ms, \mh)$, the 2D joint PDF of the red galaxies
of stellar mass $\ms$ sitting in halos of mass $\mh$, given the 2D PDF of the overall galaxy population
$p(\ms, \mh)$ inferred from Paper I,
\begin{equation}
    p\red(\ms, \mh) =  \frac{f\red(\ms, \mh)}{f\tot\red} p(\ms, \mh),
    \label{eqn:p2dred}
\end{equation}
where $f\tot\red$ is the overall red faction of all galaxies, obtained via
\begin{equation}
    f\tot\red =\iint f\red(\ms, \mh)p(\ms, \mh)\,\dd\mh\dd\ms.
\end{equation}

As described in Paper I, {\ihod} predicts the $w_p$ and $\ds$ signals for a given galaxy sample by combining
the predicted signals from individual narrow redshift slices, each of which is described by a single standard
HOD.

For deriving standard HODs within redshift slices for the red galaxies, we need
\begin{equation}
    p\red(\mh|\ms) = \frac{p\red(\mh, \ms)}{p\red(\ms)},
\end{equation}
while $p\red(\ms)$ is the predicted {\it parent}~(i.e., including observed and unobserved galaxies) SMF of the
red galaxies normalized by their total number density $n\tot\red$,
\begin{equation}
    p\red(\ms) = \frac{\phi\red(\ms)}{n\tot\red} = \int_0^{+\infty} p\red(\mh, \ms)\,\dd \mh.
\end{equation}

Finally, we arrive at the HOD of red galaxies at any redshift $z$ as
\begin{equation}
    \avg{N\red(\mh | z)} = \left(\frac{\mathrm{d}n}{\mathrm{d}\mh}\right)^{-1}\int_{\ms^{0}}^{\ms^{1}} p\red(\mh |
    \ms)\, \Phi\red_{\mathrm{obs}}(\ms|z)\,\dd\ms,
    \label{eqn:nmhihod}
\end{equation}
where $\Phi\red_{\mathrm{obs}}(\ms|z)$ is the {\it observed} SMF of red galaxies at redshift $z$, directly
accessible from the survey. For modelling the samples defined in Figure~\ref{fig:samsel} for the {\ihod}
analysis, we measure the observed galaxy SMF at each redshift, and then obtain the HOD for that redshift slice
using Equation~\eqref{eqn:nmhihod}. In this way, we avoid the need to explicitly model the sample
incompleteness as a function of $\ms$ and/or $\mh$.

For the blue galaxies, we apply the same procedures above to obtain $\avg{N\blue(\mh | z)}$ from
$p\blue(\mh, \ms)$, by substituting $f\red(\ms, \mh)$ with $f\blue(\ms, \mh){\equiv}1-f\red(\ms, \mh)$ in
equations~\ref{eqn:p2dred}$-$\ref{eqn:nmhihod}.

Fig.~\ref{fig:qihods} illustrates the two sets of coloured {\ihod}s derived from the best-fit hybrid~(top row)
and halo~(bottom row) quenching models. In each row, the left and right panels display $\lg(\dd
N(\ms|\mh)/\dd\lg\ms)$, the average log-number of galaxies per dex in stellar mass within halos at fixed mass,
for the red and blue populations, respectively. The white and black contour lines highlight the central and
satellite galaxy occupations separately on the $\ms$-$\mh$ plane. All panels reveal the same generic pattern,
consisting of a tight sequence that corresponds to the SHMR of the central galaxies, and a cloud underneath
occupied by the satellite galaxies. The level of similarity exhibited by the red galaxies is especially high
between the two quenching models~(left column).

However, comparing the left and right panels in the same row~(i.e., red vs. blue galaxies in the same
quenching model), the red centrals are more preferentially sitting in the high-$\ms$ and high-$\mh$ region
than in the low-$\ms$ and low-$\mh$ one, while the opposite is true for the blue centrals.  This segregation
happens regardless of quenching models, confirming our notion that it is difficult to unambiguously
disentangle the two quenching directions, despite their orthogonality, by merely examining the quenching trend
with $\ms$, or some surrogate of $\mh$ that has substantial scatter about the true $\mh$~(e.g., group
richness).

The satellites are quenched by $\mh$ in both models, but are also partially by $\ms$ in the hybrid model.
Thus, there are more high-$\ms$ blue satellite galaxies within massive halos in the halo quenching
model~(bottom right panel) than in the hybrid model~(top right panel). In addition, the low mass halos in the
hybrid quenching model are more likely to host blue dwarf satellites than in the halo quenching model.

The two sets of {\ihod} models, presented in this Section and in Fig.~\ref{fig:qihods}, are the analytical
foundation that allow us to predict the $w_p$ and $\ds$ signals as functions of the four parameters in each
quenching model. Any difference shown in Fig.~\ref{fig:qihods} between the two models will be propagated to
the different behaviours in the final predictions of $w_p$ and $\ds$, and is thus detectable by comparing the
two sets of predictions to the measurements from SDSS galaxies.

\section{Constraints on the Two Quenching Models}
\label{sec:constraint}

\begin{table*}
\centering \caption{Description, prior specifications, and posterior constraints of the parameters in the
halo~(top) and hyrbid~(bottom) quenching models. All the priors are uniform distributions running across the entire range of possible values for the
parameters, and the uncertainties are the $68\%$ confidence regions derived from the 1D posterior probability
distributions.}
\begin{tabular}{ccccc}
\hline
\hline
Parameter & Description & Uniform Prior Range & prior case & fixed case \\
\hline
\hline
\multicolumn{5}{|c|}{Halo Quenching Model $Q_{\mathrm{halo}}$} \\
\hline
 $\lg\mhqc\;[\hmsol]$ & Characteristic halo mass for central galaxy quenching  &[11.0, 15.5]& $12.20_{-0.08}^{+0.07}$  & $12.25_{-0.06}^{+0.05}$   \\
 $\muc$ & Pace of central galaxy quenching with halo mass            &[0.0, 3.0]  & $0.38_{-0.03}^{+0.04} $  & $0.42_{-0.03}^{+0.03} $   \\
 $\lg\mhqs\;[\hmsol]$ & Characteristic halo mass for satellite galaxy quenching&[11.0, 15.5]& $12.17_{-0.10}^{+0.12}$  & $12.30_{-0.23}^{+0.17}$   \\
 $\mus$ & Pace of satellite galaxy quenching with halo mass          &[0.0, 3.0]  & $0.15_{-0.02}^{+0.03} $  & $0.16_{-0.03}^{+0.03} $   \\
\hline
\multicolumn{5}{|c|}{Hybrid Quenching Model $Q_{\mathrm{hybrid}}$} \\
\hline
 $\lg\msq\;[\hhmsol]$ & Characteristic stellar mass for central and satellite quenching&[9.0, 12.0]&$10.50_{-0.04}^{+0.04}$  & $ 10.55_{-0.03}^{+0.03}$   \\
 $\mu$ & Pace of galaxy quenching with stellar mass                   &[0.0, 3.0]       &$0.69_{-0.06}^{+0.06} $  & $ 0.66_{-0.05}^{+0.06} $   \\
 $\lg\mhq\;[\hmsol]$ & Characteristic halo mass for satellite galaxy quenching  &[11.0, 15.5]     &$13.76_{-0.14}^{+0.15}$  & $ 13.63_{-0.11}^{+0.10}$   \\
 $\nu$ & Pace of satellite galaxy quenching with halo mass            &[0.0, 3.0]       &$0.15_{-0.05}^{+0.05} $  & $ 0.18_{-0.04}^{+0.05} $   \\
\hline
\end{tabular}
\label{tab:pripos}
\end{table*}

\begin{figure}
\begin{center}
    \includegraphics[width=0.9\columnwidth]{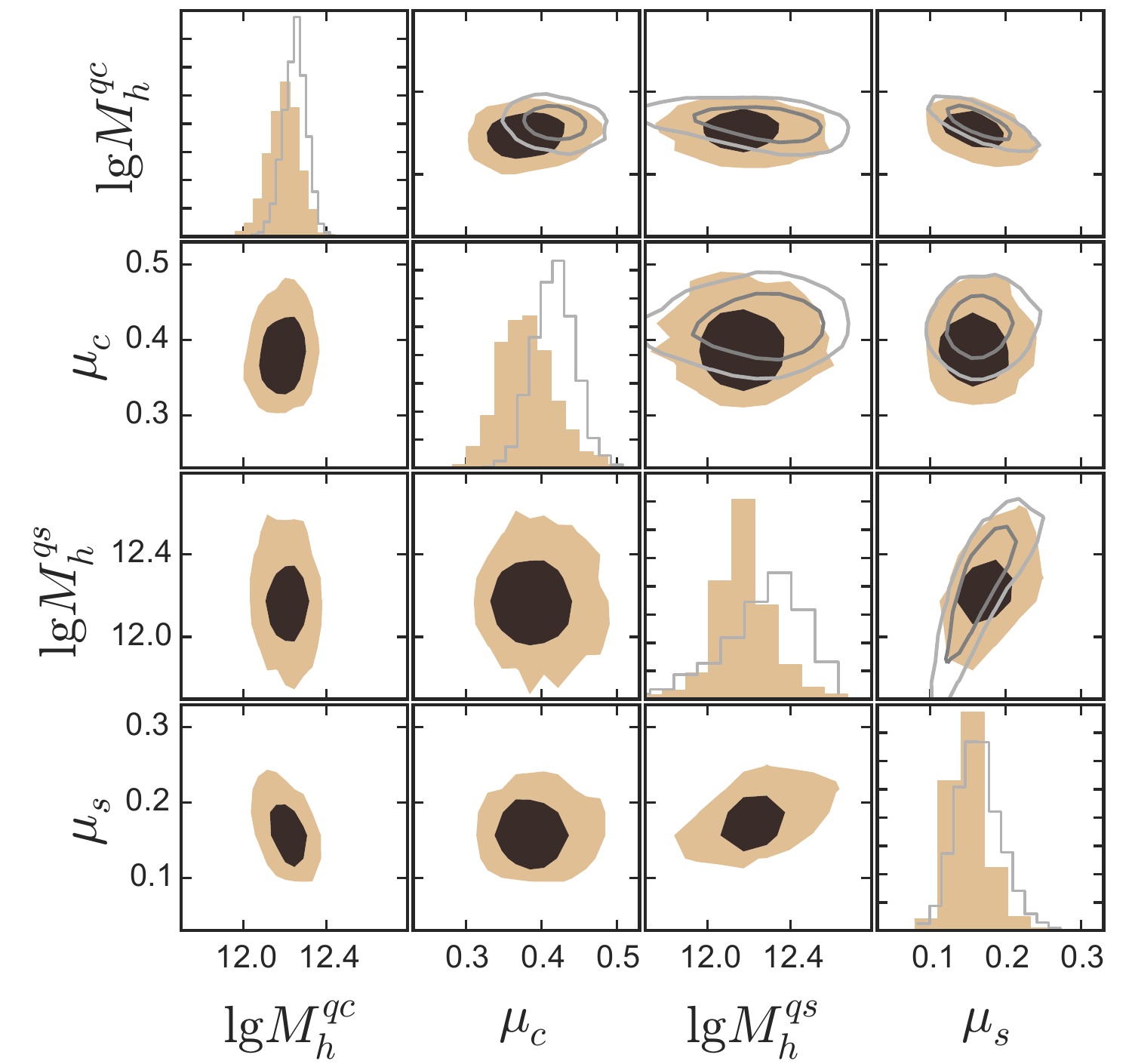}
    \caption[]{\input{figures/glory_gd15/caption.tex}
}
\end{center}
\end{figure}

\begin{figure}
\begin{center}
    \includegraphics[width=0.9\columnwidth]{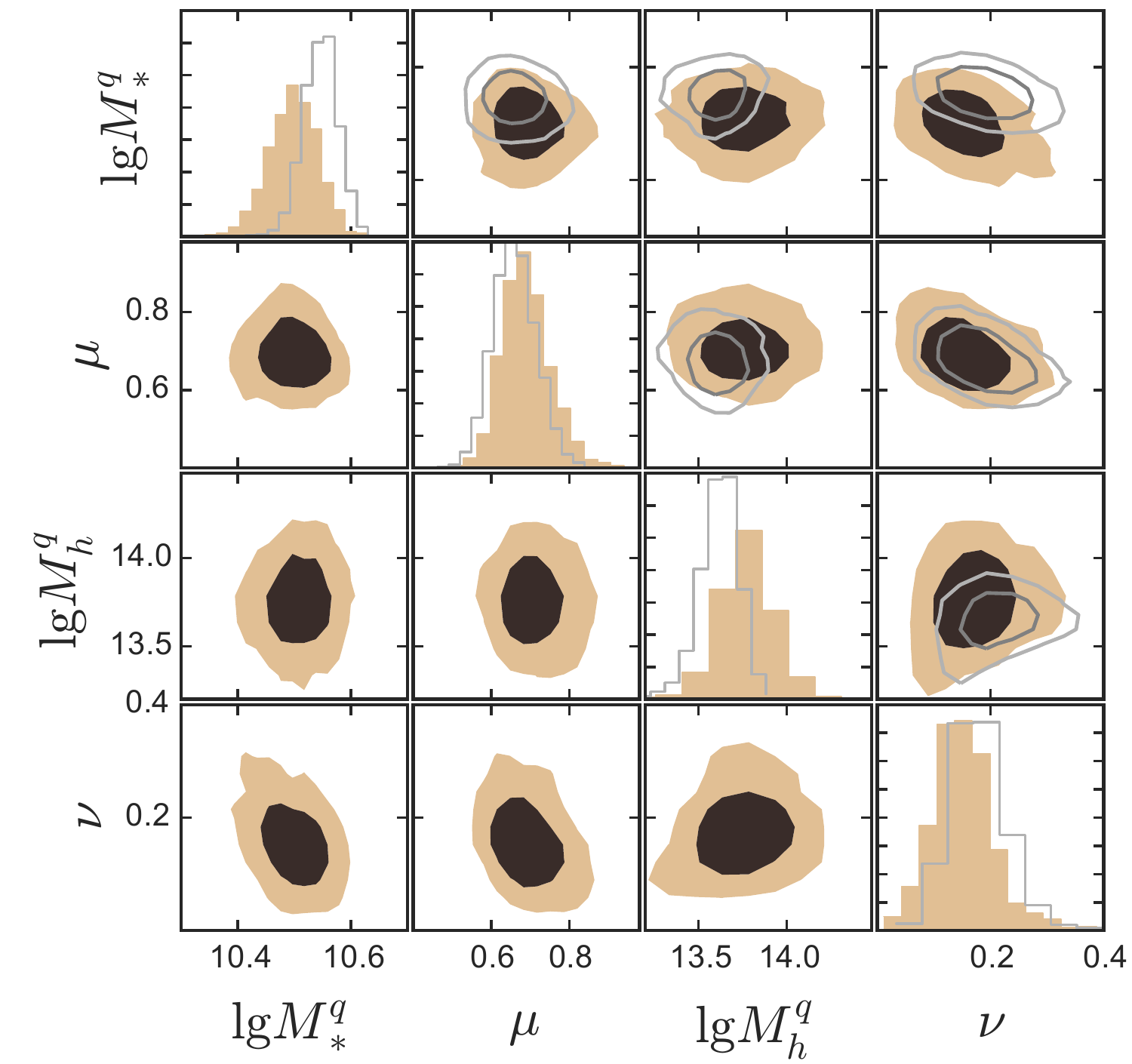}
    \caption[]{\input{figures/glory_p10/caption.tex}
}
\end{center}
\end{figure}

\begin{figure*}
\begin{center}
    \includegraphics[width=0.8\textwidth]{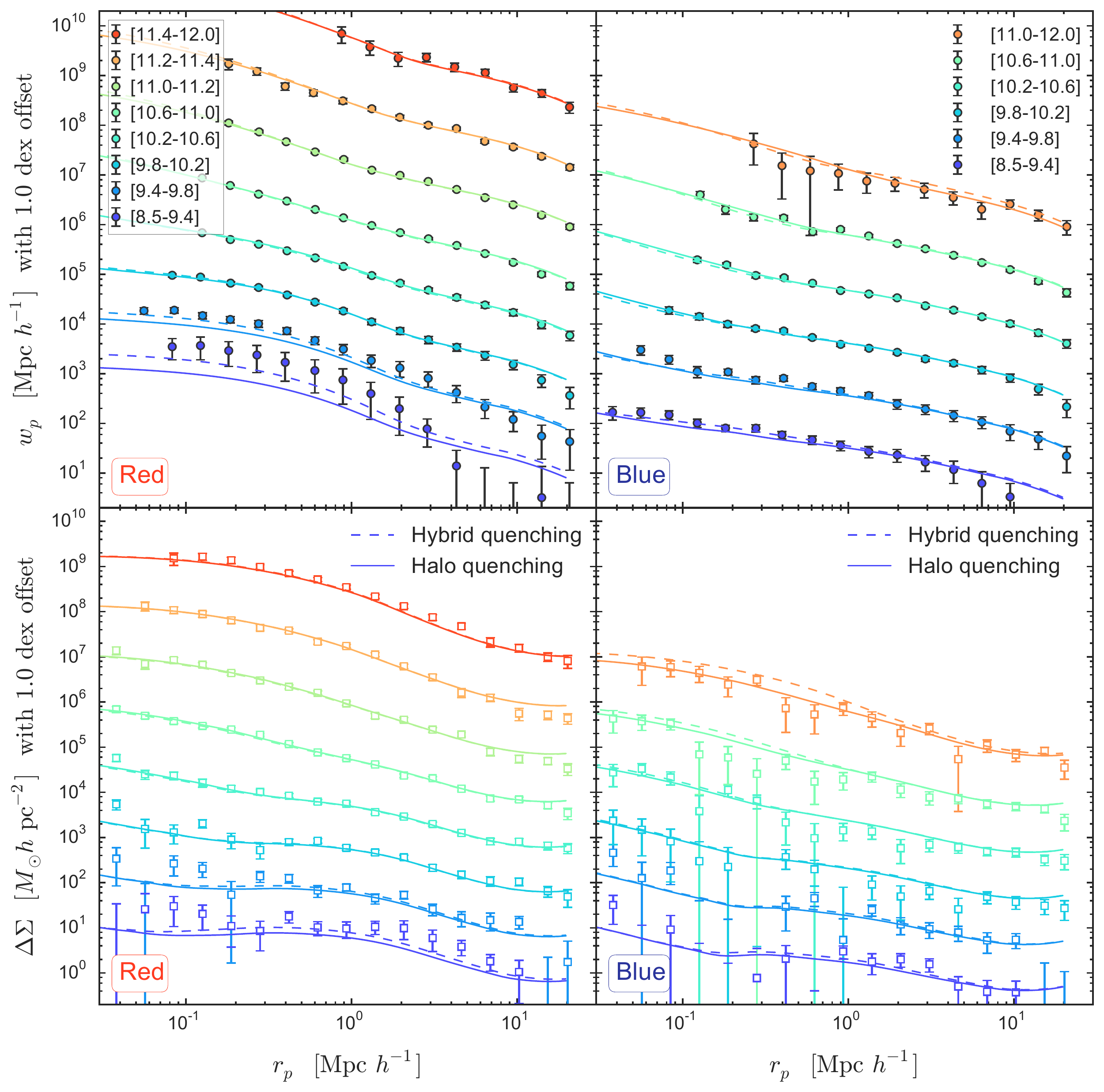}
    \caption[]{\input{figures/qbestfits/caption.tex}
}
\end{center}
\end{figure*}

\begin{figure*}
\begin{center}
    \includegraphics[width=0.8\textwidth]{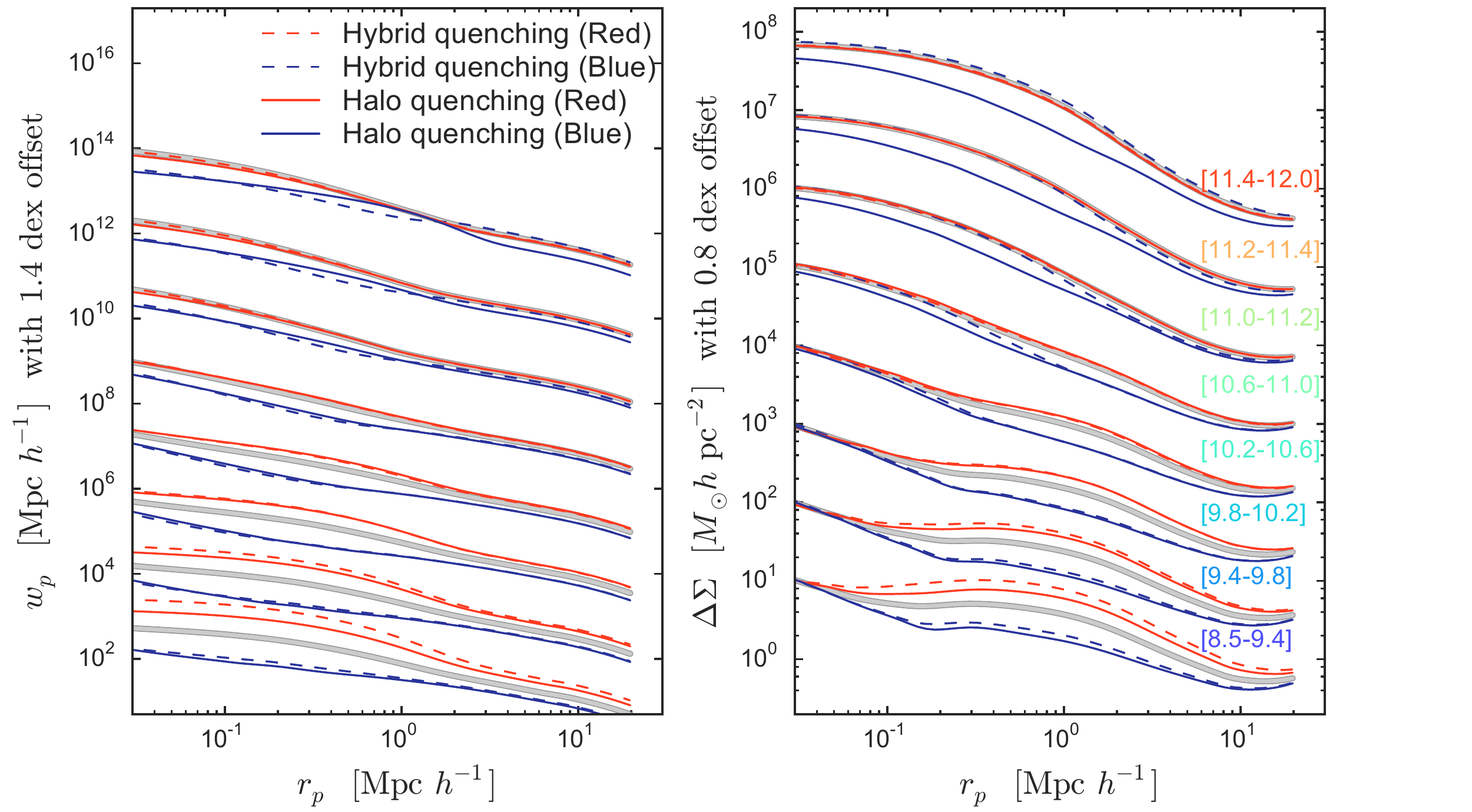}
    \caption[]{\input{figures/qcompall/caption.tex}
}
\end{center}
\end{figure*}

\begin{figure*}
\begin{center}
    \includegraphics[width=0.8\textwidth]{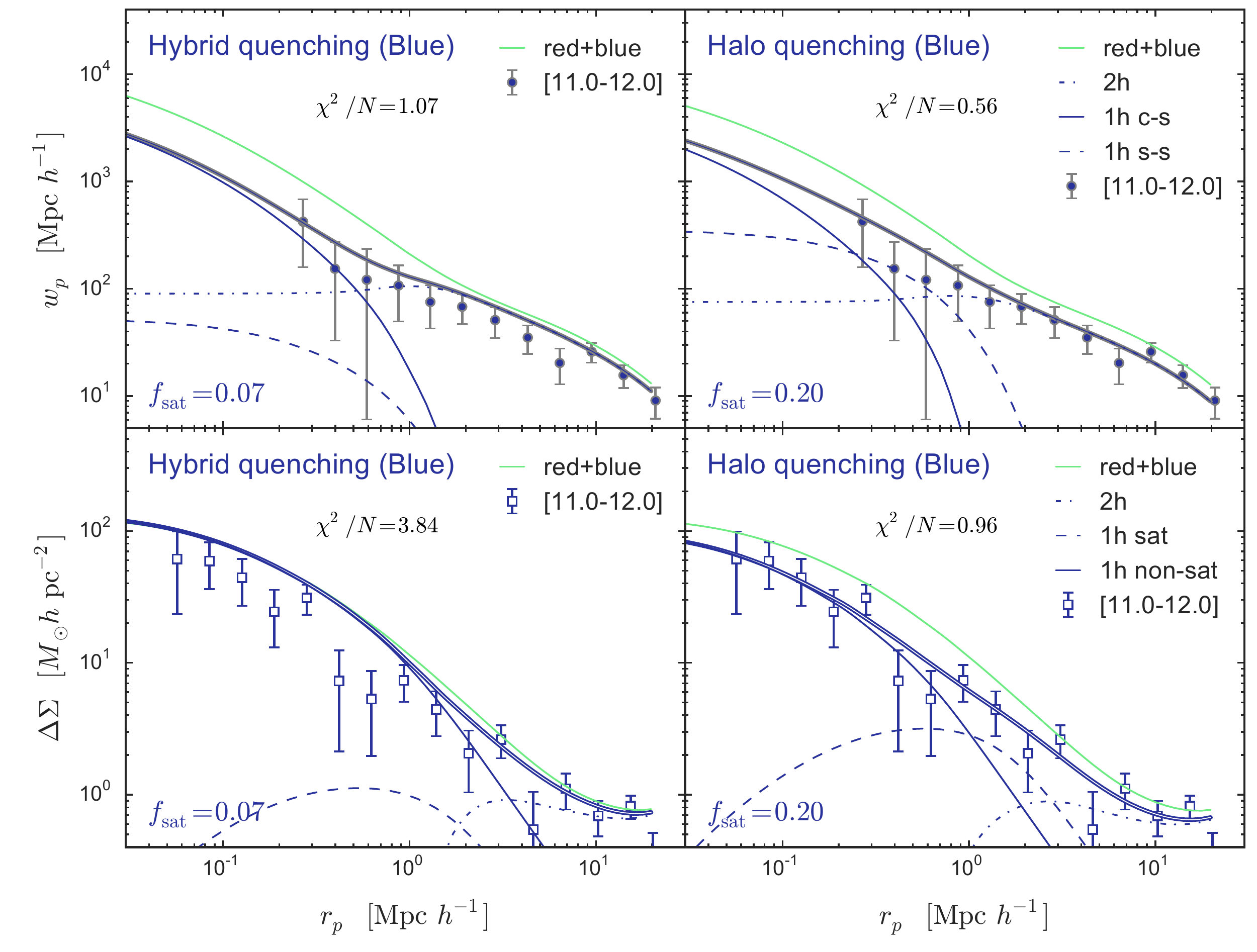}
    \caption[]{\input{figures/qdecomps_blue/caption.tex}
}
\end{center}
\end{figure*}

\subsection{Constraints of the Quenching Parameters}
\label{subsec:posterior}

Ideally one would constrain both the {\ihod} parameters and the quenching parameters together, by
simultaneously fitting to the $w_p$ and $\ds$ measurements of the overall, red, and blue galaxies. However,
since the measurements of the overall population have the highest signal-to-noise ratio and the overall
{\ihod} does not include quenching, it is conceptually more reasonable to adopt a two-step scheme. In the
first step we constrain the {\ihod} parameters using only the measurements of the galaxy samples without dividing by colour~(i.e.,
Paper I). In the second step, when constraining the quenching parameters, we either fix the best-fit
{\ihod} parameters~(i.e., the ``fixed case'') or input the {\ihod} constraints from Paper I as priors~(i.e.,
the ``prior case''). In particular, for the prior case we draw the {\ihod} parameters from the joint prior
distribution represented by the Markov Chain Monte Carlo~(MCMC) samples derived in Paper I. For each quenching
model, we adopt the results from the prior case as our fiducial constraint in the following analysis.

In addition to the powerful statistical features of the {\ihod} framework inherited from Paper I, our
quenching analysis also adds two important advantages compared to the traditional HOD modelling of red and
blue galaxies. Firstly, traditional HOD studies of red and blue galaxies treat the two populations
independently, so that the total number of HOD parameters inevitably doubles compared to the modelling of the
overall galaxy population~\citep[e.g.,][]{tinker2013, puebla2015}. In our analysis, the red and blue
populations are derived not from scratch, but by filtering the overall {\ihod} with the red fraction predicted
by each quenching model, which is described by only four simple yet physically meaningful parameters. Our
method also guarantees that the sum of the red and blue SHMRs is mathematically identical to the overall SHMR.
Secondly, the traditional method usually parameterizes the red galaxy fraction as a 1D function of halo mass,
while our method affords a 2D function of $f\red$ defined on the $\ms$--$\mh$ plane, which is crucial to the
task of examining stellar mass as a potential driver for quenching.

For each quenching model, we infer the posterior probability distributions of the four model parameters from
the $w_p$ and $\ds$ measurements of the eight red and six blue galaxy samples within a Bayesian framework. We
model the combinatorial vector $\mathbf{x}$ of the $w_p$ and the $\ds$ components of the red and blue galaxies
as a multivariate Gaussian, which is fully specified by its mean vector~($\bar{\mathbf{x}}$) and covariance
matrix~($C$). The Gaussian likelihood is thus
\begin{equation}
    \mathcal{L}(\mathbf{x} | \boldsymbol{\theta}) =
    |C|^{-1/2}\exp\left(-\frac{(\mathbf{x}-\bar{\mathbf{x}})^TC^{-1}(\mathbf{x}-\bar{\mathbf{x}})}{2}\right),
\label{eqn:gauloglike}
\end{equation}
where
\begin{equation}
   \boldsymbol{\theta} \equiv \{ \lg\msq, \mu, \lg\mhq, \nu \}
\end{equation}
in hybrid quenching, and
\begin{equation}
   \boldsymbol{\theta} \equiv \{ \lg\mhqc, \muc, \lg\mhqs, \mus \}
\end{equation}
in halo quenching.

We adopt flat priors on the model parameters, with a uniform distribution over a broad interval that covers
the entire possible range of each parameter~(see the 2nd column of Table~\ref{tab:pripos}).  The final
covariance matrix $C$ is assembled by aligning the error matrices of $w_p$ and $\ds$ measured for individual
coloured samples along the diagonal blocks of the full $N \times N$ matrix. We ignore the weak covariance
between $w_p$ and $\ds$ (with the covariance being weak due to the fact that $\ds$ is dominated by shape
noise), and between any two measurements of the same type but for different stellar mass or coloured samples.

Fig.~\ref{fig:glory_gd15} presents a summary of the inferences from the halo quenching model analysis, showing
the 1D posterior distribution for each of the four model parameters~(diagonal panels), and the $95\%$ and $68\%$
confidence regions for all the parameter pairs~(off--diagonal panels). In the panels of the lower triangle, we
highlight the results from our fiducial model, i.e., the prior case, employing the {\ihod} parameter
constraints from Paper I as priors. In each panel of the upper triangle, we compare the constraints from the
fiducial analysis~(filled contours) to that of the fixed case analysis, which keeps the {\ihod} parameters at
their best-fit values derived from Paper I. The two analyses are consistent with each other, implying that the
explanation of the red and blue signals does not require any modification in the description of the overall
galaxy population. The two inferred characteristic halo mass scales are very similar to the critical shock heating mass
scale, $\mhqc{\sim}\mhqs{\sim}M_{\mathrm{shock}}$, while the two powered-exponential indices, $\muc$ and
$\mus$, indicate that the central and the satellite quenching transition differently across that shared
characteristic halo mass. We defer the detailed discussion of the physical implications of the halo quenching
constraints in Section~\ref{sec:physics}. The $68\%$ confidence regions of the 1D posterior constraints are
listed in Table~\ref{tab:pripos}.

Similarly, Fig.~\ref{fig:glory_p10} presents the constraints on the hybrid quenching model. The critical
stellar mass for quenching all galaxies is $\msq{=}3.16(\pm{0.31})\times10^{10}\hhmsol$, echoing the
characteristic stellar mass for downsizing at the low redshift. The stellar mass quenching index $\mu$ is
slightly below unity, the value required for maintaining the observed redshift-independence of Schechter $M^*$
and faint-end slope of the star-forming galaxies in the stellar mass quenching formalism proposed in P10. The
characteristic halo mass for the quenching of satellites is much higher than $\mhqs$, albeit with a similar
quenching index of $\nu{=}0.15\pm{0.05}$.

\subsection{Best-fit Model Predictions}
\label{subsec:bestfit}

Fig.~\ref{fig:qbestfits} compares the clustering~(top row) and g-g lensing~(bottom row) signals measured from
SDSS~(points with errorbars) to those predicted by the best-fit halo~(solid lines) and hybrid~(dashed line)
quenching models, for the eight red~(left column) and the six blue~(right column) stellar mass samples.  In
terms of the overall goodness-of-fit, the best-fit halo quenching model yields a $\chi^2$ of $701.0$, while
the hybrid quenching model has a worse $\chi^2$ of $736.8$. The reduced $\chi^2$ values are thus $1.60$ and
$1.68$ for the halo and hybrid quenching, respectively, both providing reasonable fits to the data,
considering that the uncertainties in the measurements of the low-$\ms$ samples are under-estimated.  We defer
a discussion of the statistical significance of both best fits to the upcoming section.

For the red galaxy samples, the two quenching models predict very similar signals except for the two lowest
stellar mass bins. Unfortunately the $w_p$ measurements in these two bins are severely affected by the
underestimated cosmic variance due to the small volumes, with highly correlated uncertainties on all scales.
Therefore, neither quenching model gives an adequate fit to their $w_p$ signals.  The $\ds$ signals of the two
lowest mass bins are less affected by cosmic variance because the measurement error is dominated by shape
noise, and are thus better described by the two quenching model predictions.

The two quenching models also predict very similar signals for the blue galaxies, except for the high mass
ones with $\lg\ms{>}11$. While both quenching models give adequate fits to the $w_p$ signals of these massive blue
galaxies, the halo quenching model produces much better fit to their $\ds$ signals than the hybrid quenching
model, driving most of the difference in the log-likelihoods~(i.e., the $\chi^2$ values) of the two best-fit
models. This difference revealed by the massive blue galaxies, as will be discussed further later, is the key
to distinguishing the two quenching models.

Fig.~\ref{fig:qcompall} highlights the split between the red and blue galaxies from the overall population in
the $w_p$~(left) and $\ds$~(right) signals, predicted by the two best-fit quenching models for the eight
stellar mass bins marked in the right panel. In each panel, the thick gray curves are the {\ihod} predictions
for the overall galaxy samples, which bifurcate into the thin red and blue curves, i.e., predictions for the
red and blue galaxies. Solid and dashed line styles indicate the halo and hybrid quenching models,
respectively. As seen in Fig.~\ref{fig:qbestfits}, the two quenching models predict very similar bifurcation
signatures, except for the high mass bins where the hybrid quenching predicts a stronger large-scale bias, a
weaker small-scale clustering strength, but a stronger small-scale g-g lensing amplitude, than the halo
quenching for the blue galaxies.  Unfortunately the measured $w_p$ signals for the high mass galaxies are cut
off at small scales due to fibre collision, and the measurement uncertainties in the large-scale $w_p$ are not
small enough to distinguish the two quenching predictions~(top right panel of Fig.~\ref{fig:qbestfits}).

Therefore, the g-g lensing of the massive blue galaxies clearly provides the most discriminative information,
as shown in the right panel of Fig.~\ref{fig:qcompall}. For blue galaxies above $10^{11}\hhmsol$, the halo
quenching model predicts substantially lower weak lensing amplitudes than the hybrid model on all distance
scales, and thus provides a much better fit to the measurements (see bottom right panel of
Fig.~\ref{fig:qbestfits}).

To understand the discrepancy between the two quenching predictions for the massive blue galaxies, we show the
decomposition of $w_p$~(top row) and $\ds$~(bottom row) signals predicted by the best-fit hybrid~(left column)
and halo~(right column) quenching models for the $\lg\ms{=}11.0\mbox{-}12.0$ blue sample in
Fig.~\ref{fig:qdecomps_blue}.  In each panel, the data points with errorbars and the thick blue curve are the
measured and predicted signals for the blue sample, while the {\ihod} prediction for the overall sample is
shown by the thin green curve.  The best-fit quenching model prediction is then decomposed into contributions
from the 1-halo and 2-halo~(thin dotted) terms. For $w_p$ the 1-halo term includes the contributions from
centrel-satellite pairs~(thin solid; ``1-h c-s'') and satellite-satellite~(thin dashed; ``1-h s-s'')  pairs;
For $\ds$ the 1-halo term consists of a satellite term~(thin dashed) and a non-satellite term~(thin solid).
We also include a point source stellar mass term in $\ds$, which is model-independent and negligible on most
of the relevant scales~(not shown here). Most importantly, the 1-halo non-satellite term is directly related
to the average dark matter density profile of the host halos~(including both the main halos for centrals and
the subhalos for satellites), and its amplitude is proportional to the average mass of those halos.  The halo
quenching model clearly provides a much better fit to the data than the hybrid model, with factors of two and
four improvement in $\chi^2$ for $w_p$ and $\ds$, respectively. In addition, Fig.~\ref{fig:qdecomps_blue}
shows the crucial advantage of including g-g lensing in the joint analysis --- since the best-fit hybrid
quenching model adequately describes the galaxy clustering~($\chi^2/N{=}1.07$), its deficiency would not be
exposed unless we compare the $\ds$ predictions to data~($\chi^2/N{=}3.84$).

Fig.~\ref{fig:qdecomps_blue} reveals two major differences between the two quenching model predictions: 1) the
halo quenching model predicts a much higher satellite fraction among the massive blue galaxies than the hybrid
model, hence the more prominent ``1-halo satellite'' terms; and 2) the average (sub)halo mass of those massive
blue galaxies predicted by the halo quenching model is much lower compared to the hybrid model prediction,
hence the lower g-g lensing amplitudes and better fit to the data.  Roughly speaking, since the hybrid
quenching model relies on the stellar mass to quench central galaxies, it tends to place central galaxies at
fixed $\ms$ into similar halos regardless of their colours. However, in the halo quenching model any galaxies
that are unquenched have to live in lower mass halos than their quenched counterparts with similar $\ms$.  In
the section below we will argue that, the discrepancy between the average halo masses of the massive blue
galaxies predicted by the two quenching models is insensitive to the details in the model parameters,
therefore can be used as a robust feature for identifying the dominant quenching driver.

\subsection{Origin of Host Halo Mass Segregation between Red and Blue Centrals}
\label{subsec:origin}

\begin{figure*}
\begin{center}
    \includegraphics[width=0.8\textwidth]{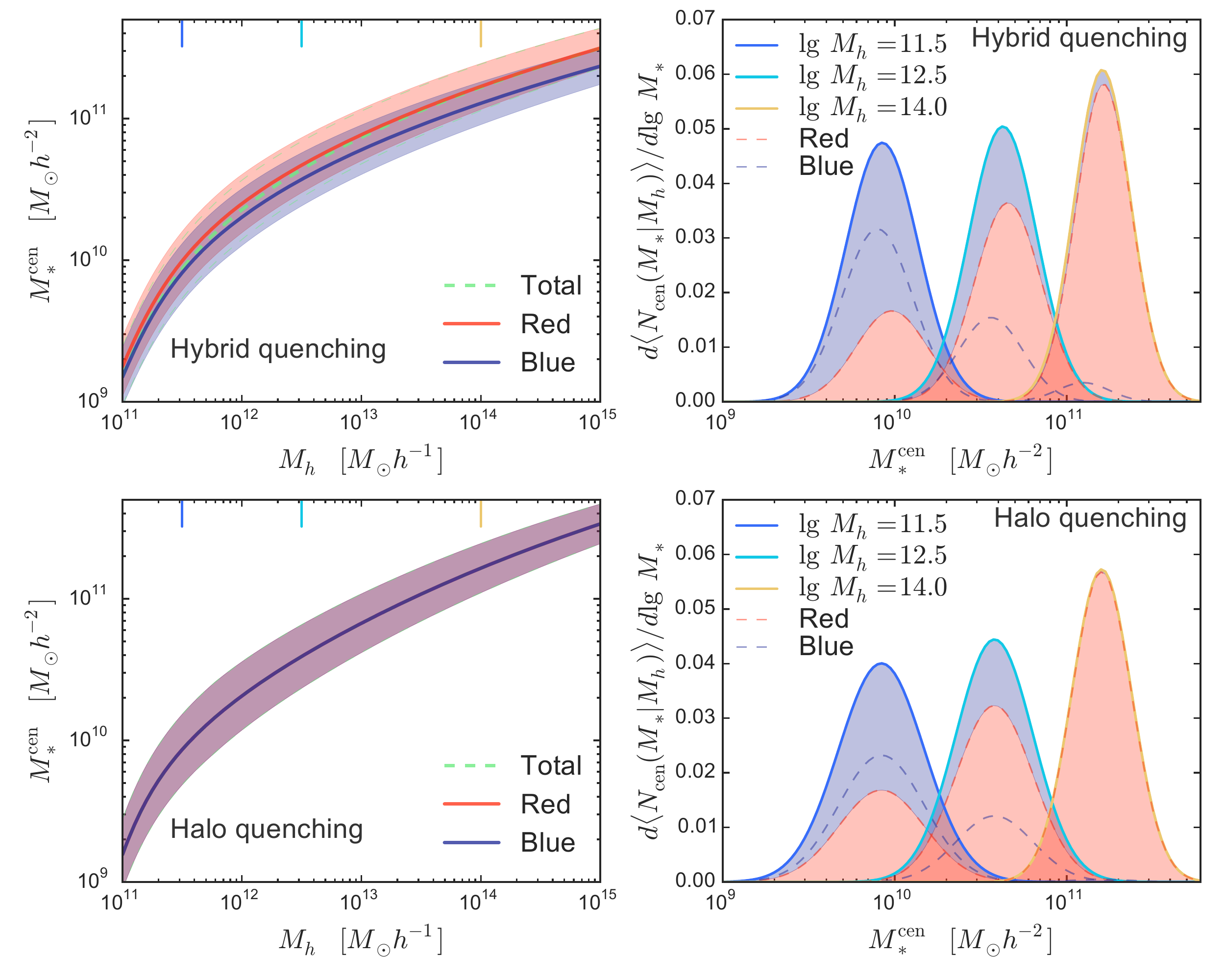}
    \caption[]{\input{figures/qshmrs/caption.tex}
}
\end{center}
\end{figure*}

Comparison between the two best-fit predictions in Section~\ref{subsec:bestfit} reveals that
$\langle\mh|\ms\rangle$, the average host halo mass at fixed stellar mass~(i.e., the mean halo-to-stellar mass
relation; HSMR), is potentially the key discriminator of the two types of quenching models. In particular, by
predicting a lower $\langle\mh|\ms\rangle$ for the blue centrals, the halo quenching model provides a much
better fit to the $w_p$ and $\ds$ signals of the massive blue galaxies than the hybrid quenching model.  But
before going any further, we need to understand the cause of this discrepancy between the two quenching
models, especially to answer the following questions. Firstly, what is the origin of the host halo mass
segregation between the two colours? Secondly, is the halo quenching necessary for predicting the strong
segregation in $\langle\mh|\ms\rangle$ between the red and blue centrals, and can the stellar mass quenching
process produce an equally low halo mass for the massive blue centrals with a different $\mu$?

For red or blue central galaxies, the conversion from the mean SHMR~(i.e., $\langle\ms|\mh\rangle$) to its
inverse relation, the HSMR, is highly non-trivial. Using the blue centrals as an example, the HSMR can be
computed from
\begin{equation}
    \langle\mh|\ms\rangle\cen\blue = \int p\cen\blue(\mh|\ms)\mh\,\dd\mh,
    \label{eqn:hsmr}
\end{equation}
where
\begin{align}
    p\cen\blue(\mh|\ms) &= \frac{p\cen\blue(\ms|\mh) p\cen\blue(\mh)}{p\cen\blue(\ms)} \nonumber\\
    &\propto p\cen\blue(\ms|\mh) f\cen\blue(\ms, \mh) \frac{\dd n(\mh) }{\dd\mh}
    \label{eqn:hsmr2}
\end{align}
In the above equation, $p\cen\blue(\ms|\mh)$ is the PDF of blue central galaxy stellar mass at fixed $\mh$,
determined by the mean SHMR of the blue centrals and its scatter, $f\cen\blue$ is the blue fraction of
centrals, and $\dd n/\dd\mh$ is the halo mass function. Therefore, for given cosmology the HSMR of the blue
central galaxies has two components, the blue central SHMR~(both mean and scatter) and the blue fraction of
centrals. To understand $\langle\mh|\ms\rangle$ for both colours more quantitatively, we start by examining
the red and blue SHMRs predicted by the two models.

The top and bottom panels in the left column of Fig.~\ref{fig:qshmrs} show the mean SHMRs of the total, red,
and blue central galaxies, predicted by the best-fit hybrid and halo quenching models, respectively.  Coloured
bands indicate the logarithmic scatters about the mean relations. The hybrid quenching model predicts a
segregation in $\ms$ between the red and blue central galaxies at fixed halo mass, as the high $\ms$ galaxies
are more likely to be quenched. The halo quenching model, however, predicts exactly the same SHMRs for all
three populations, as galaxies at fixed halo mass are equally likely to be quenched regardless of stellar
mass. The red and blue segregation in $\ms$, or the lack thereof, is best illustrated
in the two right panels of Fig.~\ref{fig:qshmrs}, using three halo masses as examples~($\lg\ms{=}11.5$, $12.5$, $14.0$).

In each panel, the total filled area for each halo shows the stellar mass distribution of central galaxies in
that halo. The width of the distribution decreases with halo mass due to the flattening of SHMR on the high
mass end. Under each distribution, the red and blue shaded areas represent the contributions from the red and
blue centrals, so that the sum of the red and blue SHMRs exactly recovers the total SHMR.  In the hybrid
quenching model for any given halo mass, the red galaxy distributions are shifted to higher $\ms$ compared to
the blue distributions, which are indicated by the dashed histograms and are equivalent to the blue shaded
regions. The halo quenching model produces zero such shift.  The non-zero shift in the hybrid model drives the
SHMR of the blue central galaxies to become shallower than that of the red centrals as seen in the top left
panel of Fig.~\ref{fig:qshmrs}. Naively, one might think that this shift will also cause the blue centrals to
reside in more massive halos than the red centrals if we simply compare the inverse functions of the two SHMRs
--- a shallower~(blue) SHMR maps the same $\ms$ on the y-axis to a higher halo mass on the x-axis.

\begin{figure*}
\begin{center}
    \includegraphics[width=0.8\textwidth]{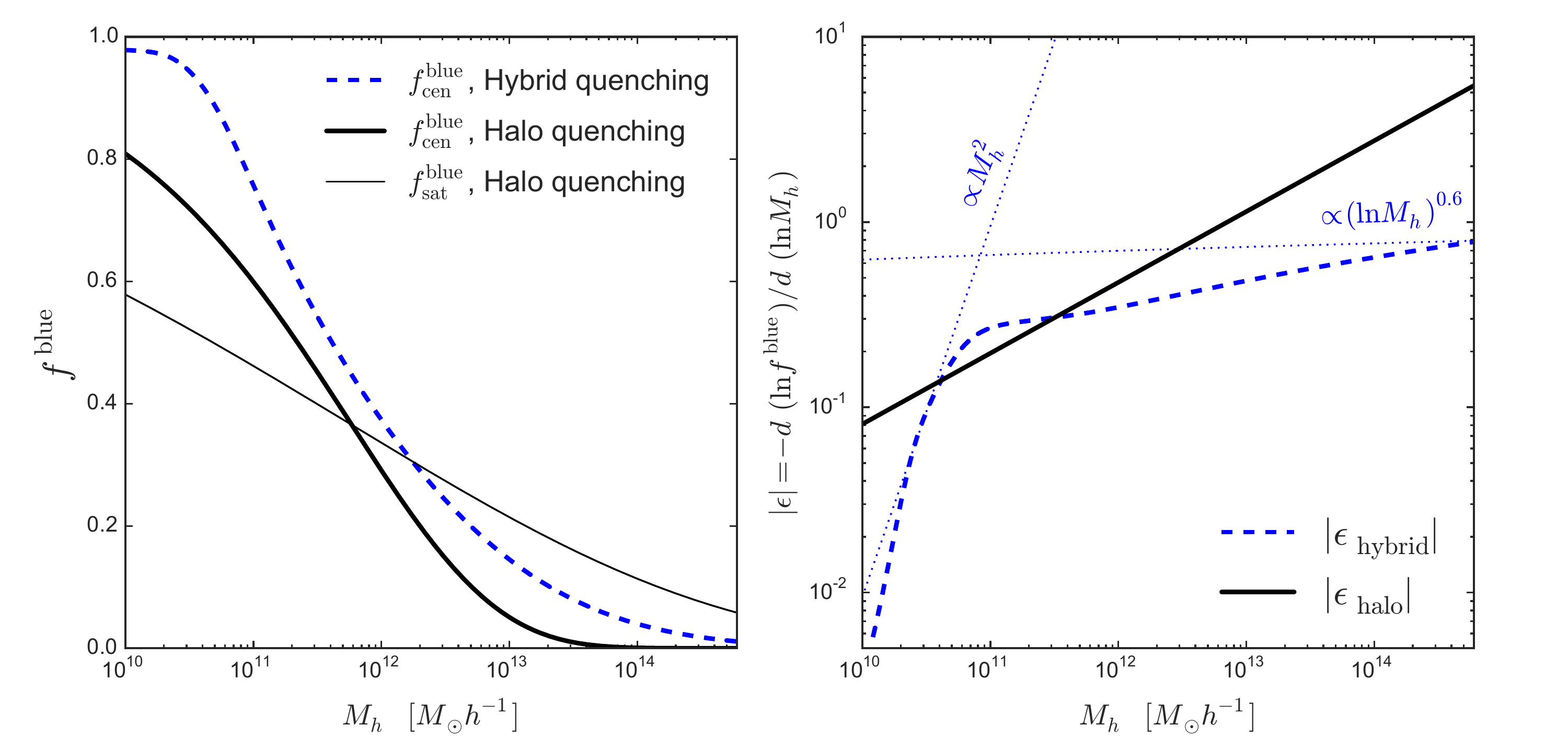}
    \caption[]{\input{figures/fblue/caption.tex}
}
\end{center}
\end{figure*}

However, a more careful inspection of the segregation patterns reveals a second, and much more important
difference in the predicted fraction of blue galaxies among centrals --- blue centrals persist in all halo
masses in the hybrid quenching model, but barely show up in the $10^{14}\hmsol$ halos in the halo quenching
model. The left panel of Fig.\ref{fig:fblue} illustrates the blue fractions as functions of $\mh$ predicted by
the best-fit halo~(thick black solid) and hybrid~(thick blue dashed) quenching models.  The amplitude of
$f\cen\blue$ in hybrid quenching also depends on $\ms$ and the blue dashed curve is the average blue fraction
over all galaxies above $10^{8}\hhmsol$.  While $f\cen\blue$ in the halo quenching case strictly follows the
powered exponential form~(i.e., equation~\ref{eqn:frcenhalo}), in the hybrid case it is affected by both the
stellar mass quenching and the slope of the SHMR.  We also show the blue fraction of satellites~(thin solid)
derived by the halo quenching analysis, which exhibits a slower decline with $\mh$ compared to that of
centrals.

The blue galaxy fractions of centrals decline rapidly with halo mass in both quenching models,  but the speed
of decline varies differently as a function of halo mass between the two models.  To investigate this
quantitatively, we define the ``central galaxy quenching rate'' as a function of halo mass, $\epsilon(\mh)$,
as the logarithmic rate at which the blue fraction declines with halo mass, $\dd\ln f\blue\cen/\dd\ln\mh$,
which is shown in the right panel of Fig.\ref{fig:fblue} for each model. As expected, the halo quenching
produces a steady increase of $|\epsilon|$ with $\mh$
\begin{equation}
    \epsilon_{\mathrm{halo}}\equiv\left(\frac{\dd \ln f\blue\cen}{\dd\ln \mh}\right)_{\mathrm{halo}} \propto -\mh^{\muc} {\simeq}
    - \mh^{0.35}.
    \label{eqn:fblueslopehalo}
\end{equation}
For the hybrid quenching case, $f\blue\cen(\mh)$ experiences a rapid decline at low $\mh$ and then a gradual
one at high $\mh$. This shift in gear can be understood as follows.  The central galaxy quenching rate
$\epsilon_{\mathrm{hybrid}}$ depends on both the $f\blue\cen(\ms)$ and the derivative of the SHMR, so that
\begin{equation}
    \epsilon_{\mathrm{hybrid}}\equiv\left(\frac{\dd \ln f\blue\cen}{\dd\ln\mh}\right)_{\mathrm{hybrid}} =\left(\frac{\dd \ln
	    f\blue\cen}{\dd\ln\ms}\right)_{\mathrm{hybrid}}
    \left(\frac{\dd\ln\ms}{\dd\ln\mh}\right).
\label{eqn:fblueslope}
\end{equation}
Since $f\blue\cen(\ms)$ also has a powered exponential form~(see equation~\ref{eqn:frcenhybrid}),
\begin{equation}
    \left(\frac{\dd \ln f\blue\cen}{\dd\ln\ms}\right)_{\mathrm{hybrid}} \propto - {\ms^{\mu}}.
    \label{eqn:fblueslopems}
\end{equation}
The slope of the SHMR $\dd\ln\ms/\dd\ln\mh$ is tightly constrained by Paper I, which described
the SHMR $f_{\mathrm{SHMR}}{\equiv}\exp\avg{\ln\ms(\mh)}$ as the inverse of
\begin{equation}
    \ln\frac{\mh}{M_1}  =  \beta \ln m + \left(\frac{m^\delta}{1 + m^{-\gamma}}- \frac{1}{2}\right) ,
    \label{eqn:shmr}
\end{equation}
where $m\equiv\ms/M_{*,0}$, $M_{*m, 0}{\sim}2\times10^{10}\hhmsol$ and $M_1{\sim}1.3\times10^{12}\hmsol$ are
the characteristic stellar and halo mass that separate the behaviours in the low and high mass ends, and the
remaining parameters control the running slopes of the SHMR. Assuming reasonable values of the slope
parameters~(i.e., $\beta{\sim}0.33$, $\delta{\sim}0.42$, $\gamma{\sim}1.21$; see Paper I),
equation~\ref{eqn:shmr} can be approximated by
\begin{equation}
    \ln\frac{\mh}{M_1} \simeq \left\{ \begin{array}{ll}
            \beta\ln m - 0.5,&\mbox{ $m \ll 1$} \\
        m^{\delta} + [\beta\ln m - 0.5] ,&\mbox{ $m \gg 1$}.
\end{array} \right.
\end{equation}
Clearly, the SHMR is a steep power-law relation at the low mass end, with $\ms \propto \mh^{1/\beta} \sim
\mh^3$, whereas at the high mass end the slope of SHMR is very shallow, with $\ms \propto (\ln\mh)^{1/\delta}
\sim (\ln\mh)^{2.4}$.

Therefore, the slope of the SHMR is
\begin{equation}
    \frac{\dd\ln\ms}{\dd\ln\mh} \simeq \left\{ \begin{array}{ll}
            1/\beta,&\mbox{ $\mh \ll M_1$} \\
        (\delta\ln\mh)^{-1} ,&\mbox{ $\mh\gg M_1$}.
\end{array} \right.
\label{eqn:shmrslope}
\end{equation}
Combining equations~\ref{eqn:fblueslope}, \ref{eqn:fblueslopems}, and \ref{eqn:shmrslope}, we arrive at
\begin{equation}
     % \left(\frac{\dd \ln f\blue\cen}{\dd\ln\mh}\right)_{\mathrm{hybrid}}
     \epsilon_{\mathrm{hybrid}}
    \propto \left\{ \begin{array}{ll}
            -\mh^{\mu/\beta},&\mbox{ $\mh \ll M_1$} \\
         -(\ln\mh)^{(\mu-\delta)/\delta} ,&\mbox{ $\mh\gg M_1$}.
\end{array} \right.
\label{eqn:fblueslopefinal}
\end{equation}
Assuming $\mu=0.67$ from the best-fit hybrid quenching model, we have
\begin{equation}
     % \left(\frac{\dd \ln f\blue\cen}{\dd\ln\mh}\right)_{\mathrm{hybrid}}
     \epsilon_{\mathrm{hybrid}}
    \propto \left\{ \begin{array}{ll}
              -\mh^{2}   ,&\mbox{ $\mh \ll M_1$} \\
          -(\ln\mh)^{0.6}      ,&\mbox{ $\mh\gg M_1$}.
\end{array} \right.
\label{eqn:fblueslopefinal2}
\end{equation}
The above equation is shown as the dotted lines on the right panel of Fig.~\ref{fig:fblue}, roughly
reproducing the two distinctive asymptotic behaviours of $\epsilon_{\mathrm{hybrid}}$ at the low and high mass
ends. The actual slope of $\epsilon_{\mathrm{hybrid}}$ is steeper than predicted by
equation~\ref{eqn:fblueslopefinal2} at high masses, where equation~\ref{eqn:shmrslope} becomes less accurate.

The comparison between $\epsilon_{\mathrm{halo}}$ and $\epsilon_{\mathrm{hybrid}}$ in
Fig.~\ref{fig:fblue}~(i.e., equations~\ref{eqn:fblueslopefinal2} and ~\ref{eqn:fblueslopehalo}) clearly
reveals that, the halo quenching model does not quench central galaxies in the low mass halos as efficiently
as the hybrid model, but by maintaining a steady quenching rate at $\epsilon(\mh)\sim-\mh^{0.35}$ the halo
quenching model is able to quench almost all centrals in the very massive halos. The hybrid quenching model,
however, is relatively inefficient to quench massive central galaxies in the very high mass halos. When
calculating the HSMR using equation~\ref{eqn:hsmr2}, this difference in $\epsilon$ completely dominates the
effect due to the slight difference between the two coloured SHMRs.  Therefore, the stellar mass quenching,
due to its slow central galaxy quenching rate on the high mass end, is incapable of producing a strong
segregation in the HSMR between the two colours. In order for the hybrid quenching model to mimic the steeper
slope of $\epsilon_{\mathrm{halo}}(\mh)$, the stellar mass quenching trend would have to drop so precipitously
that the abundance of blue galaxies is cut off beyond some maximum stellar mass, which is ruled out by the
observed SMFs of blue galaxies~(see Fig.~\ref{fig:smf_gd15}). Therefore, we further emphasize that this slow
quenching rate with halo mass in the hybrid model is caused by the changing slope of SHMR across $M_1$, and is
thus insensitive to the stellar mass quenching prescriptions, e.g., the value of $\mu$.

To summarize the findings above using the quenching diagram of Fig.~\ref{fig:qmodels}, the steep
slope~($M_*{\sim}M_h^3$) of the SHMR below $M_1$ makes the SHMR more aligned with the quenching arrow along
the $\ms$-axis~(i.e., stellar mass quenching, see top left panel of Fig.~\ref{fig:qmodels}), causing
progressively more galaxies to be quenched at higher halo mass. Above $M_1$, however, the SHMR becomes
shallower~($M_*{\sim}(\ln M_h)^{2.4}$) and is almost perpendicular to the quenching arrow, leaving a
substantial number of blue centrals in massive halos.  As a result, the massive blue centrals are extremely
scarce in the $10^{14}\hmsol$ halos in the halo quenching model~(bottom right panel of Fig.~\ref{fig:qshmrs}),
but have a much stronger presence within those halos in the hybrid quenching model~(top right panel of
Fig.~\ref{fig:qshmrs}). By the same token, a strong segregation in the host halo mass between the red and blue
centrals would points to the necessity of a dominant halo mass quenching for the central galaxies.

% all results come in here.

\section{Comparing the Hybrid and Halo Quenching Models}
\label{sec:result}

\begin{figure*}
\begin{center}
    \includegraphics[width=1.0\textwidth]{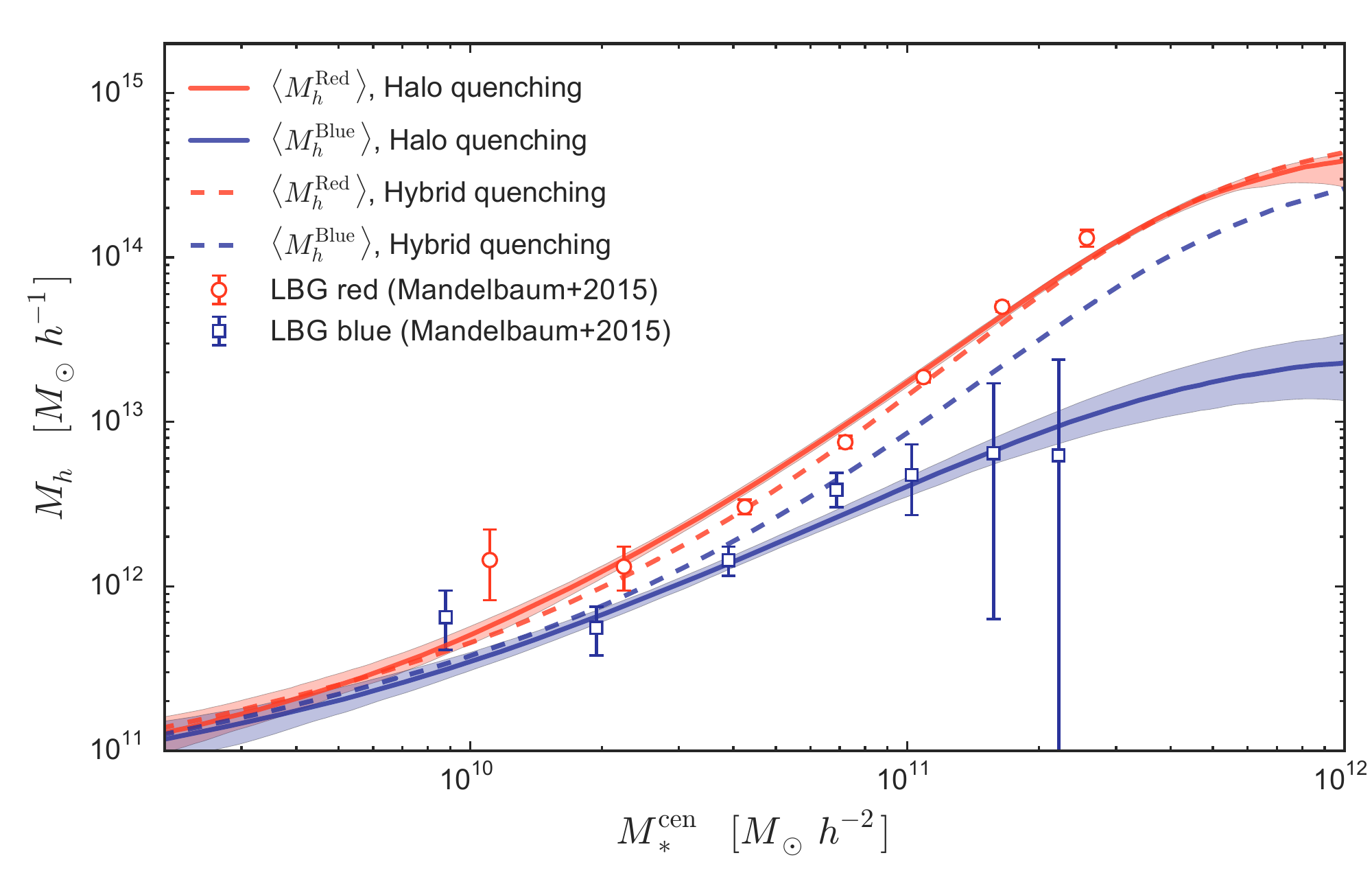}
    \caption[]{\input{figures/qhsmrs/caption.tex}
}
\end{center}
\end{figure*}

In this section we perform a robust comparison between the two quenching models in two ways, an internal one
based on Bayesian Information Criterion~(BIC) described in Section~\ref{subsec:bic}, and an external one based
on cross-validation~(Section~\ref{subsec:mbg}), which is motivated by the quenching impact on the average halo
mass of the massive blue galaxies quantitatively explained in Section~\ref{subsec:origin}.

\subsection{Internal Model Comparison: Bayesian Information Criterion}
\label{subsec:bic}

In Bayesian applications, pairwise comparisons between models $M_1$ and $M_2$ are often based on the Bayes
factor $B_{12}$, which is defined as the ratio of the posterior odds, $P(M_1|\mathbf{x})/P(M_2|\mathbf{x})$,
to the prior odds, $\pi(M_1)/\pi(M_2)$. In our case, the bayes factor is
\begin{equation}
B_{12} =
\frac{P(Q_{\mathrm{halo}}|\mathbf{x})}{P(Q_{\mathrm{hybrid}}|\mathbf{x})}\frac{\pi(Q_{\mathrm{hybrid}})}{\pi(Q_{\mathrm{halo}})},
\end{equation}
so that a $B_{12}$ above unity indicates the data favor halo quenching and a $B_{12}$ below points to hybrid
quenching. However, in most practical settings~(as is the case here) the prior odds are hard to set precisely,
and model selection based on BIC is widely employed as a rough equivalent to selection based on Bayes factors.
The BIC~(a.k.a., ``Schwarz information criterion''), is defined as
\begin{equation}
\mathrm{BIC} = -2 \ln \mathcal{L}_{\mathrm{max}} + k \ln n,
\label{eqn:bic}
\end{equation}
where $\ln \mathcal{L}_{\mathrm{max}}$ is the maximum likelihood value, $k$ is the number of parameters, and
$n$ is the number of data points. \citet{kass95} argued that in the limit of large $n$~($n{=}439$ in our
analyses),
\begin{equation}
    \frac{-2\ln B_{12} - (\mathrm{BIC}_{\mathrm{halo}} - \mathrm{BIC}_{\mathrm{hybrid}})}{-2\ln B_{12}} \longrightarrow 0
    \label{eqn:b12}
\end{equation}
i.e., $\Delta\mathrm{BIC}=\mathrm{BIC}_{\mathrm{halo}} - \mathrm{BIC}_{\mathrm{hybrid}}$ can be viewed as a
rough approximation to $-2\ln B_{12}$, so that $\Delta\mathrm{BIC}{<}0(<-10)$ indicates that
$Q_{\mathrm{halo}}$ is favored~(strongly) and  $\Delta\mathrm{BIC}{>}0(>10)$ points (strongly) to
$Q_{\mathrm{hybrid}}$.

The $\Delta\mathrm{BIC}$ between the two quenching models is $-35.8$, which corresponds to an asymptotic value
of $\ln B_{12}{=}17.9$ according to equation~\ref{eqn:b12}. Therefore, based on the BIC test, the clustering
and the g-g lensing measurements of the red and blue galaxies strongly favor the halo quenching model against
the hybrid quenching model, and the halo mass is the more statistically dominant driver of galaxy quenching
than stellar mass.

The two quenching models are non-nested models with the same $k$ and $n$, so the second term of
equation~\ref{eqn:bic} that penalizes model complexities is the same in both quenching models. The BIC test is
then equivalent to the alternative Akaike information criterion that is based on relative likelihoods, or a
simple $\Delta\chi^2$ test~(i.e., $\Delta\chi^2{=}35.8$). These tests all point to the halo
mass as the main driver of quenching.
% 439
% -817.453756 hybrid chi-square = 370.4 * 2
% -797.856536 halo chi-square = 350.81 * 2
% det
% -447.05

\subsection{External Model Comparison: Halo Masses of Massive Blue Centrals}
\label{subsec:mbg}

The discussion in Section~\ref{subsec:origin} points to a potentially smoking-gun test of the two quenching
models, by comparing the host halo mass of the massive red and blue central galaxies predicted from the two
best-fit quenching models, to other mass measurements for observed groups/clusters with red and blue centrals
within the same redshift range.  Unfortunately, clusters with blue centrals are systematically under-selected
by most of the photometric cluster finders based on matching to the red sequence, while spectroscopic group
catalogues constructed from friends-of-friends algorithms do not have large enough volume for finding many
massive clusters.

Recently, \citet{mandelbaum2015} constructed a sample of locally brightest galaxies~(LBGs) from the SDSS main
galaxy sample, by adopting a set of isolation criteria carefully calibrated against semi-analytic mock galaxy
catalogues to minimize the satellite contamination rate~\citep{wang2015}. The resulting LBG sample is thus a
subset of all massive central galaxies, but with excellent purity of central galaxy membership and zero bias
against blue colour.  Therefore, the LBGs are ideal for our purpose of measuring the segregation in halo mass
between the red and blue centrals.

\citet{mandelbaum2015} measured the average host halo mass of the LBGs directly by fitting an NFW density
profile~(after projection to 2D) to the weak lensing signals measured below $1\hmpc$. Fig.~\ref{fig:qhsmrs}
compares the host halo mass measured as a function of LBG stellar mass~(data points with errorbars) to that
predicted by the best-fit halo~(solid) and hybrid~(dashed) quenching models. The errorbars on the LBG
measurements are the 1-$\sigma$ uncertainties on the mean halo mass, derived from $1000$ bootstrap-resampled
datasets. The coloured bands about the solid curves are the uncertainties on the mean halo mass predicted
from the $68\%$ confidence regions. To avoid clutter, we do not show the uncertainties on the hybrid quenching
predictions, which are comparable to the halo quenching uncertainties.  The average halo mass predicted by the
halo quenching model is in excellent agreement with the measurements from the LBG sample, while the hybrid
quenching model, as expected, grossly over-predicts the halo mass for the massive blue galaxies.

The difference in the host halo mass between red and blue centrals, as predicted by the hybrid quenching
model, can be understood simply as the outcome of a larger differential growth between dark matter and stellar
mass in quiescent systems than in star-forming ones. More specifically, in quenched systems the dark matter
halos usually continued to grow after the shutdown of stellar mass growth, while in star-forming systems the
two often grew in sync, creating a bimodality in host halo mass between two colours without invoking halo
quenching.  However, this differential growth effect causes at most a factor of two difference in the average
host halo masses, too small to explain the factor of several difference observed at the high mass
end~\citep{quadri2015}.

The LBG experiment in Fig.~\ref{fig:qhsmrs} further demonstrates that the halo quenching model, employing halo
mass as the driver for galaxy quenching, is superior to the hybrid quenching model, which relies on stellar
mass to quench central galaxies.  As we explained in Section~\ref{subsec:origin}, the deficiency of the hybrid
model in describing the signals of the massive blue galaxies is intrinsic to the stellar mass quenching
mechanism, which fails to explain the rare occurrence of blue centrals in massive clusters. The combined
evidence from the BIC model comparison and the LBG experiment strongly suggests that the halo mass is the main
driver for quenching the galaxies observed in SDSS.

\section{Physical Implications of the Constraints on Halo Quenching Model}
\label{sec:physics}

\begin{figure*}
\begin{center}
    \includegraphics[width=1.0\textwidth]{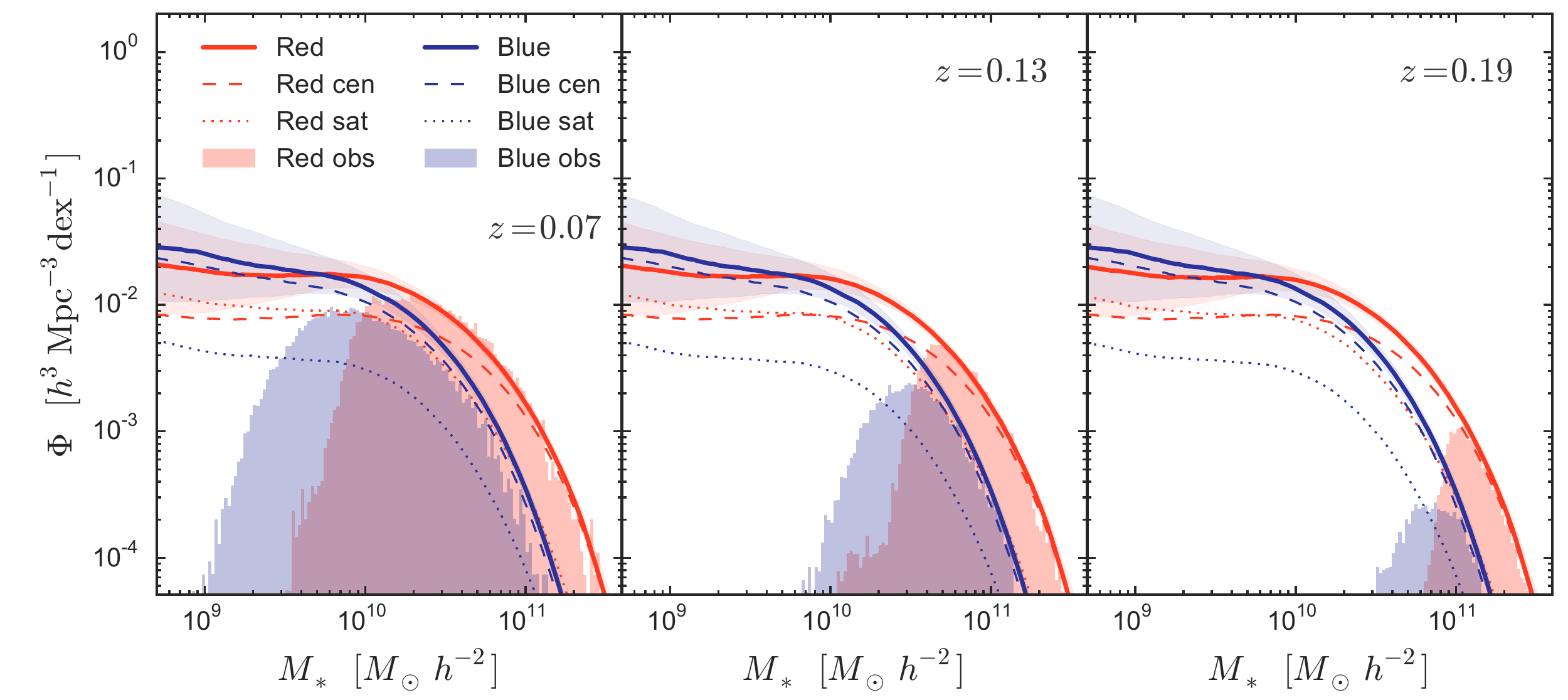}
    \caption[]{\input{figures/smf_gd15/caption.tex}
}
\end{center}
\end{figure*}

With the halo quenching model being established as the more viable scenario, we now focus back on the
physical implications of our constraints on halo quenching.

\subsection{Uniform Characteristic Halo Masses for Quenching Centrals and Satellites}
\label{subsec:mshock}

Although the halo quenching formula for centrals and satellites are decoupled in the analysis, our fiducial
constraint nonetheless recovers two very similar characteristic halo masses~($\mhqc$ and $\mhqs$) for both
species at around $1.5\times 10^{12}\hmsol$. It is very tempting to associate this uniform quenching mass
scale for both central and satellites to $M_{\mathrm{shock}}$, the critical halo mass responsible for the
turning-on of shock heating. Analytical calculations and hydrodynamic simulations both favor a
$M_{\mathrm{shock}}$ of $\sim$ few times $10^{12}\hmsol$~\citep{birnboim2003, keres2005, dekel2006}, providing
one of the most plausible explanation for the similar values of $\mhqc$ and $\mhqs$ derived statistically in
our analyses.

Conservatively speaking, even if the similarity between our inferred characteristic halo masses and
$M_{\mathrm{shock}}$ were coincidental, the consistency between $\mhqc$ and $\mhqs$ still indicates that the
quenching of centrals and satellites are somewhat coupled, most likely driven by processes that are both tied
to the potential well of the halos. For instance, the supermassive black holes~(SMBHs) could provide the
thermal or mechanical feedback required to stop the halo gas from cooling and feeding the
satellites~\citep{dimatteo2005, croton2006, somerville2008}, while regulating the growth of the central
galaxies~\citep{ferrarese2000, gebhardt2000, tremaine2002}.  \citet{hopkins2007} suggested that the SMBH mass
is largely determined by the depth of the potential well in the central regions of the system, which precedes
the assembly of halo mass, i.e., the maximum circular velocity is already half the present-day value by the
time the halo has accreted only two per cent of its final mass~\citep{bosch2014}.

\subsection{Implications for Satellite Quenching and Galactic Conformity}
\label{subsec:conformity}

The halo quenching of satellites has a slower transition across $10^{12}\hmsol$ than that of the central
galaxies~(left panel of Fig.~\ref{fig:fblue}). This rules out the possibility that halo quenching does not
distinguish between centrals and satellites.  As mentioned in the introduction, even in the hot halo quenching
scenario where gas cooling of centrals and satellites were equally inhibited, the satellites might experience
significant delays in their quenching, due to a shorter exposure to the hot halo and/or a spell of star
formation from pockets of cold gas they carried across the virial radius of the larger, hotter
halo~\citep{simha2009, wetzel2012, wetzel2013, bosch2008}. Additionally, recent observations suggest that
other processes like pre-processing during infall~\citep{haines2015}, strangulation~\citep{peng2015}, and ram
pressure stripping~\citep{muzzin2014} are all at play, contributing to the satellite quenching trend with halo
mass. The inefficiency of satellite quenching is also seen in dwarf galaxies below the stellar mass scale we
probed here~\citep{wheeler2014}.

Another interesting aspect of the halo quenching scenario is that it may help explain galactic conformity,
i.e., the observed correlation between colours of the central galaxies and their surrounding
satellites~\citep{weinmann2006, knobel2015}, because the quiescent pairs of centrals and satellites are
quenched by the common halos they share. However, this halo quenching-induced conformity only occurs among
central-satellite pairs at fixed $\ms$ of the centrals. To explain the galactic conformity observed at fixed
$\mh$, there either has to be a substantial scatter between observed $\mh$ and true $\mh$, or a secondary
process that couples the quenching of centrals and satellites within the same halo.  For instance, halos
formed earlier with higher concentration may be more likely to host quenched pairs of centrals and satellites
than their younger and less concentrated counterparts at the same $\mh$~\citep{paranjape2015}. Galactic
conformity does not appear when the centrals and satellites were quenched independently, e.g., in the hybrid
quenching scenario.

The combination of this intra-halo conformity and the correlation between clustering bias and halo mass, could
potentially explain the inter-halo conformity observed in the galaxy marked correlation
statistics~\citep{skibba2006, cohn2014} and hinted by galaxy pairs in the local volume~\citep[$z{<}0.03$;
see][]{kauffmann2013}, although a secondary quenching induced by either formation time or halo concentration
at fixed $\mh$ may be required~\citep{paranjape2015}. We will explore the conformity prediction of the halo
quenching model in the upcoming third paper of this series.

\subsection{Colour-segregated Stellar Mass Functions of Central and Satellite Galaxies}
\label{subsec:smf}

\begin{figure*}
\begin{center}
    \includegraphics[width=0.8\textwidth]{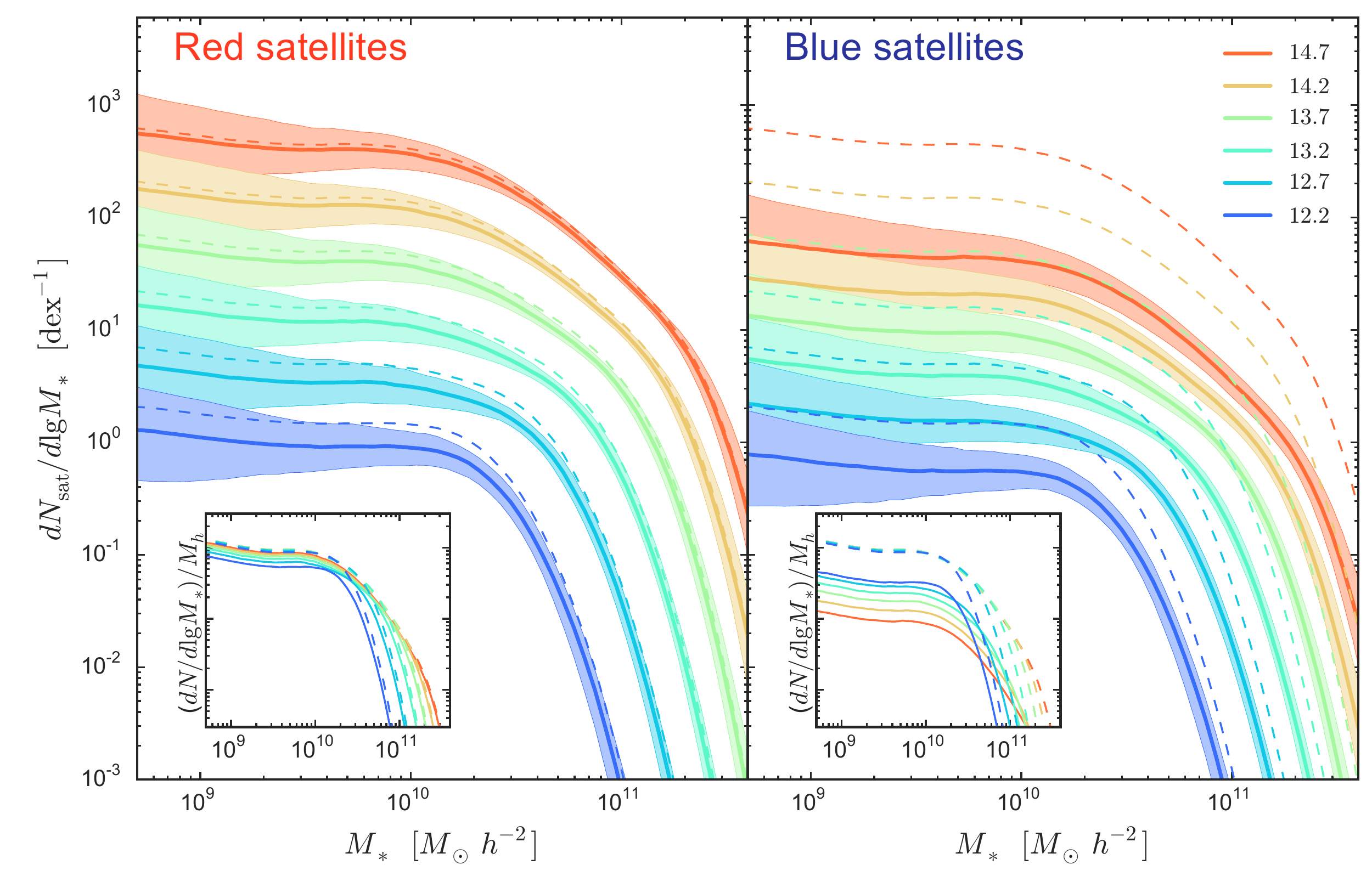}
    \caption[]{\input{figures/csmf_gd15/caption.tex}
}
\end{center}
\end{figure*}

As another consistency-check of our analysis, Fig.\ref{fig:smf_gd15} shows the underlying SMFs of the red and
blue galaxies predicted by our best-fit halo quenching model at three redshifts~($z{=}0.07$, $0.13$, $0.19$).
In each panel, the shaded bands are the 1-$\sigma$ uncertainties on the predicted SMFs, and the shaded
histograms show the observed SMFs of each colour, i.e., direct galaxy number counts at each redshift without
the $1/V_{\mathrm{max}}$ weighting.  The SMF of each colour is further decomposed into contributions from the
central~(dashed) and satellite~(dotted) galaxies.  The observed SMF at $z{=}0.07$~(left panel) is more
incomplete at the high $\ms$ end due to photometric confusions about bright sources in SDSS~(see Paper I for
details), therefore lying further below our predictions than that at the two higher redshifts.  Since the
observed SMFs are not used as input data to the constraints, the excellent agreement between our predictions
and the direct number counts on the high stellar mass end is very encouraging --- it demonstrates the great
consistency between the three key observables~(i.e., galaxy clustering, g-g lensing, and the SMFs) measured
for the two coloured populations and that predicted by the best-fit halo quenching model within the {\ihod}
framework. Furthermore, the successful recovery of the red and blue SMFs means that the halo quenching model
naturally recovers the stellar mass quenching trend observed in P10.

The best-fit halo quenching also provides concrete prediction for the conditional SMF of the red and blue
satellite galaxies, shown on the left and right panels in Fig.\ref{fig:csmf_gd15}, respectively.  The
conditional SMF is defined as the average number of satellites per dex in stellar mass at fixed halo mass,
$\langle\dd N\sat/\dd \lg \ms|\mh\rangle$, the integration of which over $\ms$ gives the commonly-used
satellite HOD, i.e., the average number of galaxies per halo above some stellar mass limit $\langle
N\sat(\ms{>}\ms^{\mathrm{lim}}) | \mh\rangle$.  In each panel, the solid curves are the red/blue satellite
SMFs within halos of six different masses ( ${\geqslant}\mhqs$), with their 1-$\sigma$ uncertainties indicated
by the shaded bands. The dashed curves are the same in both panels, showing the two-colour combined satellite
SMFs. Each inset panel shows the same set of curves as in the main panel, but each normalized by the
corresponding halo mass.  Overall, the satellite population above $\mhqs$ is dominated by the red galaxies,
and the number of blue satellite galaxies per halo mass decreases with increasing halo mass due to
progressively stronger halo quenching effect, while the total number of satellites per halo mass remains
roughly constant.

\section{Comparison to Previous Works}
\label{sec:compare}

\begin{figure*}
\begin{center}
    \includegraphics[width=1.0\textwidth]{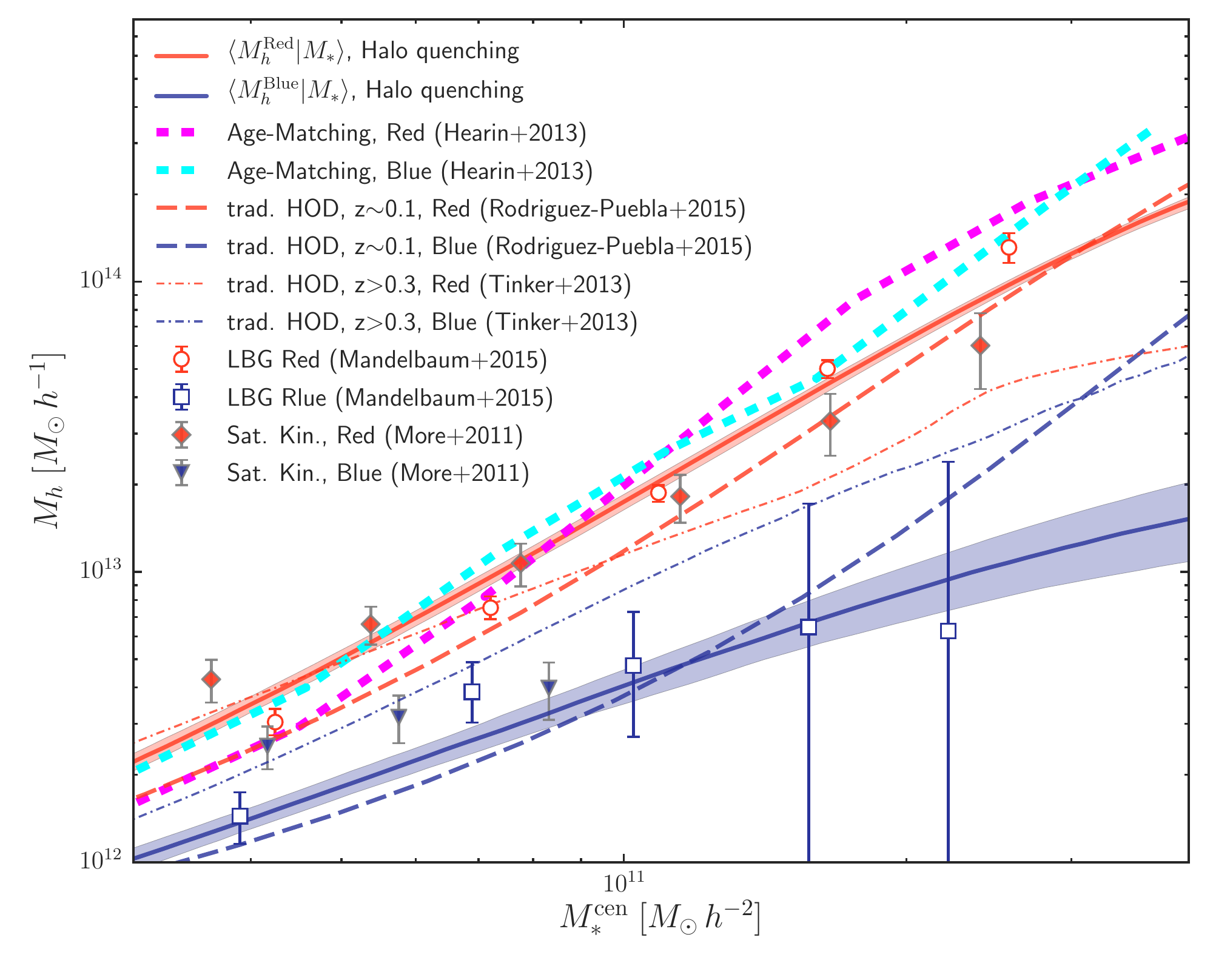}
    \caption[]{\input{figures/hsmr_comp/caption.tex}
}
\end{center}
\end{figure*}

Our quenching model is fundamentally different from previous studies of the link between galaxy colours and
the underlying dark matter halos. Here we compare our best-fit quenching model to the two main alternative
methods.  One is the separate red and blue galaxy modelling using traditional HOD
methods~(Section~\ref{subsec:sephod}), and the other is based on a modified abundance matching scheme, i.e.,
the age-matching model~(Section~\ref{subsec:age}). We summarize the comparison between our result and the
previous studies in Fig.~\ref{fig:hsmr_comp}, which zooms in on the stellar mass range of
Fig.~\ref{fig:qhsmrs} that has the maximum model discriminating power.  In addition to the LBG weak lensing
masses shown in Fig.~\ref{fig:qhsmrs}, we also include the average halo mass of the red and blue centrals
measured from satellite kinematics by~\citet{more2011}.  We note that although the various constraints and
measurements shown in Fig.~\ref{fig:qhsmrs} assumed slightly different cosmologies, the uncertainties due to
cosmology are usually much smaller than the statistical errors, and the strong bimodality~(or the lack
thereof) in the host halo mass between red and blue is independent of any changes in cosmology.  We will come
back to Fig.~\ref{fig:hsmr_comp} frequently and discuss individual comparisons in detail below.

\subsection{Comparison to Traditional HOD Models}
\label{subsec:sephod}

The most straightforward way to model the red and blue split of galaxy observables traditionally is to infer
the HODs of the overall and the red galaxies first, and subtract the two to derive the HOD of the blue.  This
approach guarantees the consistency between the three sets of HODs, but lacks the flexibility in the treatment
of the red fraction for describing the full ranges of behaviours seen in the data~\citep{zehavi2005,
zehavi2011}. A more comprehensive method is to treat the two colours separately, by prescribing independent
HODs for the two and an 1D overall quenched fraction as a function of halo mass, as done recently in
\citet{tinker2013} and \citet{puebla2015}.

As mentioned in the introduction, there are two main differences between our approach and the methods of
\citet{tinker2013} and \citet{puebla2015}.  Firstly, our quenching analysis employs only four more parameters
to explain the split into red and blue galaxies, while the traditional methods require doubling of the number
of parameters used for the overall population~(e.g., $23$ parameters in the \citealt{puebla2015} analysis, and
$27$ in \citealt{tinker2013}). Our four parameters are also more physically meaningful because they can be
directly related to the average quenching action, as we discussed in Section~\ref{sec:physics}. Secondly, our
quenching model describes the bimodality of galaxy occupation statistics in a mathematically consistent manner
--- the overall galaxy HOD is recovered by summing the inferred red and blue HODs. The other two methods do
not benefit from this consistency, making the connection between the colour-segregated HODs and the overall
HOD hard to interpret. Naively one might think that our model is a subset of the traditional HOD models that
appear more flexible by fitting red and blue separately. However, since the combination of the two separate
coloured SHMRs derived in the traditional methods usually does not fit the overall galaxy population while our
{\ihod} quenching models do, the two methods are fundamentally different descriptions of the red and blue
galaxy populations.

\begin{figure*}
\begin{center}
    \includegraphics[width=1.0\textwidth]{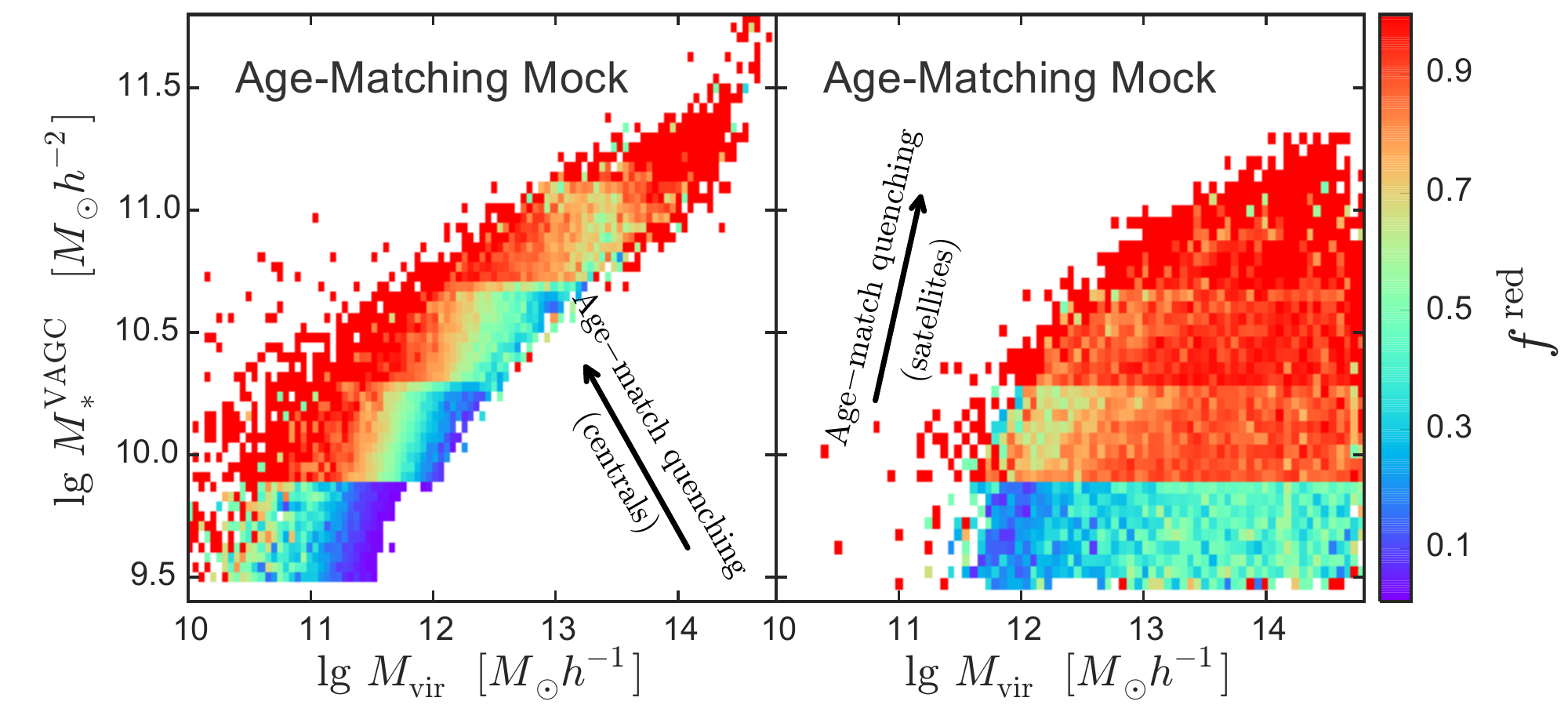}
    \caption[]{\input{figures/qmodel_cam/caption.tex}
}
\end{center}
\end{figure*}

It is worth nothing that the constraints drawn from traditional HOD modelling are unaffected by its
incapability of including stellar mass quenching, which is subdominant compared to halo mass quenching.
However, as we discussed in Section~\ref{subsec:conformity}, the halo quenching of satellites has a much
slower transition across the critical halo mass than that of centrals. Therefore, the lack of separate
treatments for the red fractions in centrals and satellites remains an important issue in those traditional
HOD models.

Using the SMFs, galaxy clustering, and g-g lensing within the COSMOS survey, \citet{tinker2013} derived the
SHMRs of active and quiescent galaxies over the redshift range between $0.2$ and $1.0$. The active/quiescent
classification is based on their separation on the optical vs. near-IR colour space, which is better at
distinguishing dusty and SF objects than using the optical colours alone. Employing the global HOD framework
of \citet{leauthaud2012}~\footnote{The parameterisation of the {\ihod} framework in Paper I is also heavily
based on the \citet{leauthaud2012} framework, but is fundamentally different in the treatment of sample
completeness and signal calculation; see Paper I for a detailed comparison between the two frameworks.}, they
applied independent central galaxy SHMR and satellite HODs to the two colours, and assumed a non-parametric
form for the red fraction as a function of halo mass using as a spline-interpolated function through five
pivotal halo masses.  For the lowest redshift bin in their analysis~($z{\sim}0.36$), the derived red and blue
SHMRs are very similar below $\mh{\sim}10^{13}\hmsol$, but strongly diverge on the high $\mh$ end, with the
red centrals having a lower average $\ms$ than the blue ones at fixed $\mh$. This divergence is equivalent to
having a stellar mass quenching at fixed halo mass, albeit in the opposite direction of the stellar mass
quenching trend observed in P10. The $\langle\mh|\ms\rangle$ relation reveals a similar trend that our halo
quenching model predicts, with the red central galaxies residing in more massive halos than the blue centrals,
but the predicted difference between the red and blue amplitudes is much smaller~(dot-dashed curves).

\citet{puebla2015} derived the separate SHMRS of red and blue galaxies using the combination of galaxy
clustering and SMFs in SDSS. Instead of using the g-g lensing as an input, they employed the SMFs measured for
the centrals and satellites separately within each colour, using the SDSS group catalogue constructed by
\citet{yang2012}. Unlike \citet{tinker2013}, they assumed a parametric form for the red fraction as a function
of halo mass. The long-dashed curves in Fig.~\ref{fig:hsmr_comp} are the $\langle\mh|\ms\rangle$ relations
inferred by \citet{puebla2015}, showing good agreement with our predictions as well as the measurements from
satellite kinematics and LBG weak lensing. The higher amplitude of $\langle\mh|\ms\rangle$ on $\ms{>}2\times
10^{11}\hhmsol$ is driven by the extrapolation of the parametric relation from lower $\ms$ rather than data.
Although \citeauthor{puebla2015}'s and our analyses make use of the same galaxies in SDSS,  the two
constraints are derived using independent methods and different measurements.  Hence, the good degree of
consistency between these results is non-trivial.

\subsection{Comparison to the Age-Matching Model}
\label{subsec:age}

An alternative way to describe the colour dependence of galaxy clustering and g-g lensing is to extend the
SHAM model to allow a secondary matching between galaxy colour and subhalo formation time, i.e., the so-called
``age-matching'' model~\citep{hearin2013}. In particular, after the usual mapping between subhalo
mass~\citep[using $v_{\mathrm{peak}}$ as a proxy; see][]{reddick2013} and galaxy stellar mass, the
age-matching method rank-orders the characteristic redshift $z_{\mathrm{starve}}$ of those subhalos at fixed
$\ms$ and matches the galaxies hosted by older subhalos to redder colours, while maintaining the observed
colour distribution of galaxies at that $\ms$. For most of the centrals~(below $10^{11}\hhmsol$),
$z_{\mathrm{starve}}$ is equivalent to the formation redshift of the subhalos $z_{\mathrm{form}}$, but at very
high $\ms$ it is dominated by $z_{\mathrm{char}}$, the first epoch at which halo mass exceeds $10^{12}\hmsol$.
The age-matching method roughly reproduces the colour and stellar mass dependences of the clustering and g-g
lensing signals~\citep{hearin2014}.

In the language of statistical quenching, the age-matching method describes the red fraction as
\begin{equation}
f\red\cen(\ms, \mh) = f\red\cen(\ms) \times m(z_{\mathrm{starve}}(\mh)|\ms),
\label{eqn:fredage}
\end{equation}
where $m(z_{\mathrm{starve}}|\ms)$ is determined by the matching between $z_{\mathrm{starve}}$ and colour at
fixed $\ms$, and $z_{\mathrm{starve}}(\mh)$ is the formation time vs. halo mass relation.  Therefore, the
age-matching process to first order assumes a stellar mass quenching, as the colour-matching step is done in
bins of $\ms$ regardless of halo mass~(the first term on the RHS of equation~\ref{eqn:fredage}), while the
secondary quenching is via formation time~(the second term on the RHS of equation~\ref{eqn:fredage}). The
combined quenching effect is best illustrated in Fig.~\ref{fig:qmodel_cam}, where we show the distribution of
red fraction on the $\ms$-$\mh$ diagram for centrals~(left) and satellites~(right), calculated from the
age-matching mock catalogue generated by~\citet{hearin2014}~(with the original NYU-VAGC stellar mass and halo
virial mass).  For the centrals, the stellar mass quenching along the vertical axis dominates, albeit in a
discrete fashion due to the binning artefact. At fixed stellar mass, since halo age is a decreasing function
of halo mass~\citep[i.e., $\dd z_{\mathrm{starve}}/\dd\mh{<}0$; c.f., figure 12 in ][]{wechsler2002}, the
secondary quenching direction of the age-matching method is a {\it reversed} halo quenching, with bluer
centrals occupying younger, thus more massive halos.  Therefore, according to the discussion in
Section~\ref{subsec:origin}, we anticipate the ``age-quenching'' to predict $\langle\mh|\ms\rangle$ relations
that are similar to the hybrid quenching model, i.e., a weak segregation in the host halo mass between the red
and blue centrals.

The magenta and cyan dotted curves in Fig.~\ref{fig:hsmr_comp} indicate the $\langle\mh|\ms\rangle$ relations
of the red and blue galaxies, measured from the age-matching mock catalogue produced
by~\citet{hearin2014}~(with both $\ms$ and $\mh$ converted to our mass units and definitions).  Compared to
the halo mass measured from satellite kinematics and LBG weak lensing, the age-matching mock heavily
over-predicts the amplitude of the relation for the blue galaxies, making the red and blue centrals occupy
halos of similar mass, despite the systematic difference in their formation times. The tiny difference in the
average halo mass between the two colours~(below $0.2$ dex over all mass scales) also switches sign across
$\ms{\sim}10^{11}\hhmsol$, where the indicator for $z_{\mathrm{starve}}$ switches from $z_{\mathrm{form}}$ to
$z_{\mathrm{char}}$. In particular, below $10^{11}\hhmsol$, age-matching predicts that the blue centrals live
in more massive halos than the red centrals, due to the anti-correlation between $z_{\mathrm{form}}$ and halo
mass --- at any given $\ms$ the younger halos that are assigned bluer centrals are also more massive. However,
this increase of halo mass with bluer colour of the centrals is in the opposite direction compared to the
observations. Above $10^{11}\hhmsol$, the characteristic redshift $z_{\mathrm{starve}}$ is dominated by
$z_{\mathrm{char}}$, which is more positively correlated with halo mass and assigns redder centrals to more
massive systems.  The overall disagreement between the $\langle\mh|\ms\rangle$ predicted by the age-matching
method and that measured from satellite kinematics and LBG weak lensing indicates that the stellar mass
quenching assumed in age-matching is not adequate, and the secondary formation-time quenching {\it at fixed
stellar mass} is strongly disfavored by the observations.

Another difference between the age-matching model and the quenching models considered in this paper is that,
by choosing formation time as a quenching indicator, the age-matching model exhibits the maximum level of
galactic assembly bias~\citep{zentner2014}, which is absent in our quenching models, by construction. However, while the ``halo
assembly bias'', namely, the dependence of halo properties on the formation history, is clearly detected in
cosmological
% simulations~\citep{sheth2004, gao2005, wechsler2006, harker2006, jing2007, hahn2007, li2008, faltenbacher2010,
simulations~\citep{sheth2004, gao2005, wechsler2006, harker2006, zhu2006, jing2007, hahn2007, croft2012}, whether it
left a significant imprint on the observed galaxy properties is still in debate~\citep{berlind2006,
    blanton2007, wang2013, miyatake2015, lin2015}. In mock galaxy catalogues constructed from
semi-analytical models and hydrodynamic simulations, galaxy clustering exhibits an assembly bias of at most
$\sim 10$ per cent~\citep{yoo2006, croton2007, zu2008, mehta2014}.\footnote{Some hydrodynamic simulations
predict much higher assembly bias effect on clustering~(${\sim}20\%$), which cannot be captured by
any current abundance matching models~\citep{chaves-montero2015}.} The great success of the halo quenching
model in quantitatively explaining the clustering and weak lensing of the red and blue galaxies, while
simultaneously recovering their respective SMFs and average halo masses, strongly suggests that the halo
quenching is the dominant process in shaping the distribution of galaxy colours observed in SDSS, and that any
impact of the galactic assembly bias should be a secondary effect, in the form of, e.g., a formation-time~(or
concentration) quenching {\it at fixed halo mass} proposed by \citet{paranjape2015}.

Finally, the physical interpretation of the halo quenching model~(as discussed in Section~\ref{subsec:mshock})
relates the quenching of galaxies to the capability of the host halos to either heat the incoming gas or keep
the hot gas from cooling, and the sharp transition of this capability across some critical halo mass is the
key to explain the strong bimodality in galaxy colours. This physical picture is fundamentally different from
the physical motivation of age-matching, in which the galaxy quenching is strictly tied to the dark matter
accretion history of halos at fixed $\ms$. However, the average halo accretion history is a smooth function of cosmic time,
therefore showing no bimodality in the formation time of halos~\citep{zhao2009}. In addition, the connection
between dark matter accretion and SFR is very complex.  Using a suite of high-resolution hydrodynamic
simulations, \citet{faucher2011} showed that the cold gas accretion rate, which is more directly related to
star formation, is in general not a simple universal factor of the dark matter accretion rate, and that
baryonic feedback can cause SFRs to deviate significantly from the external gas accretion rates.

\section{Conclusions}
\label{sec:conclusion}

We develop a novel method to identify the dominant driver of galaxy quenching in the local Universe, using the
galaxy clustering and g-g lensing of red and blue galaxies observed in SDSS. The method extends the
powerful {\ihod} framework developed in \citet{zm15} by introducing two quenching models: 1) a {\it halo}
quenching model in which the average probability of a galaxy being quenched depends solely on the main halo
mass, but in separate manners for centrals and satellites; and 2) a {\it hybrid} quenching model in which the
quenching probability of all galaxies depends on their stellar mass, with the satellite having an extra
dependence on the host halo mass.
% a halo quenching model that
% relies solely on halo mass to statistically quench both the central and satellite galaxies, and a hybrid model that assumes
% stellar mass is the main quenching driver with extra satellite quenching happening in massive halos.
% The two models quench galaxies along orthogonal directions on the $\ms$-$\mh$ plane~(Fig.~\ref{fig:qmodels}),
% resulting in distinctive patterns in which the red and blue galaxies populate massive
% halos~(Fig.~\ref{fig:qihods}).

The two quenching models predict distinctive 2D distributions of red galaxy fractions on the
$\ms$-$\mh$ plane, resulting in different patterns through which red and blue galaxies populate dark matter
halos. We then predict the clustering and g-g lensing signals of the red and blue galaxies from these two
galaxy occupation patterns using the {\ihod} framework and compare them to the measurements from SDSS. Most
importantly, the flexibility of {\ihod} allows us to
include ${\sim}80\%$ more galaxies in the analysis than the traditional HOD method, greatly enhancing our capability of
statistically distinguishing the two quenching models.

We find that the halo quenching model provides better descriptions of the bimodality in the clustering and
lensing of observed galaxies than the hybrid quenching models, mainly due to the
significantly improved fit to massive blue galaxies. We further identify that the
average host halo mass of the massive blue centrals provides the most discriminating power in testing viable
quenching models --- models with halo mass quenching generally predict a much stronger
segregation in the average host halo mass~($\langle\mh|\ms\rangle$) between the red and blue than the ones
without~(Fig.~\ref{fig:qhsmrs}).

Therefore, by comparing the $\langle\mh|\ms\rangle$ predicted by various quenching models, including the
age-matching model, to that measured directly from the satellite kinematics of galaxy clusters and the weak
lensing of locally brightest galaxies, we confirm that the best-fit halo quenching model provides excellent
agreement with the two observational measurements, while models that rely on stellar mass~(e.g., the hybrid
quenching model and the age-matching method) fail to predict the halo mass of the blue central
galaxies~(Fig.~\ref{fig:hsmr_comp}).  Furthermore, the formation time-quenching at fixed $\ms$
prescribed in the age-matching method creates a {\it reversed} halo quenching trend, therefore placing blue
and red centrals of the same $\ms$ into higher and lower mass halos, respectively. This trend is strongly
disfavored by the observations where at any given $\ms$ redder centrals on average occupy more massive
halos. Our findings indicate that any viable abundance matching scheme for assigning galaxy colours has to
reproduce the observed strong bimodality in host halo mass between red and blue centrals, especially at the
high mass end.

The derived characteristic halo masses of the central and satellite quenching have very similar values around
$1.5\times10^{12}\hmsol$, suggesting that a uniform halo quenching process is operating on both the centrals
and satellites. The derived characteristic halo mass can be interpreted by the canonical halo quenching
theory, which predicts a critical halo mass of $M_{\mathrm{shock}}{\sim}10^{12}\hmsol$.  Above this critical
mass, the virial shock is able to prevent star formation by heating the infalling gas to high temperatures,
whereas below $M_{\mathrm{shock}}$ the halo quenching rapidly turns off, creating strong bimodality in both
the colour and the spatial distribution of galaxies. The pace at which the halo mass quenching operates,
however, appears to be faster for the centrals than for the satellite galaxies, which are quenched in a more
delayed and prolonged fashion.

In the future, we anticipate that the {\ihod} halo quenching model, which accurately explains the spatial
clustering, g-g lensing, and SMFs of the red and blue galaxies, will provide an
important baseline model for explaining an even wider range of observed galaxy properties with ever-growing
precision. In the near term, i.e., the upcoming Paper III in this series,  we plan to generate realistic
colour-segregated galaxy mock catalogues using the constraints inferred from the halo quenching model analysis
in this paper, and make a comprehensive comparison with the observed galaxies to look for potential signatures
of any secondary quenching processes, e.g., due to formation time~\citep{paranjape2015} and galaxy
compactness~\citep{woo2015} at fixed halo mass.

\section*{Acknowlegements}

We thank Aldo Rodr\'iguez-Puebla, Surhud More, and Jeremy Tinker for kindly providing their measurements.  We
also thank Hung-Jin Huang for useful discussions.  We thank David Weinberg and Zheng Zheng for carefully
reading an earlier version of the manuscript and for giving detailed comments and suggestions that have
greatly improved the manuscript.  YZ and RM acknowledge the support by the Department of Energy Early Career
Program, and the Alfred P. Sloan Fellowship program.

%%%%%%%%%%%%%%%%%%%%%%%%%%%%%%%%%%%%%%%%%%%%%%%%%%



\bibliographystyle{mnras}
\bibliography{biblio} %

% \begin{thebibliography}{99}
% \bibitem[\protect\citeauthoryear{Author}{2012}]{Author2012}
% Author A.~N., 2013, Journal of Improbable Astronomy, 1, 1
% \bibitem[\protect\citeauthoryear{Others}{2013}]{Others2013}
% Others S., 2012, Journal of Interesting Stuff, 17, 198
% \end{thebibliography}

%%%%%%%%%%%%%%%%%%%%%%%%%%%%%%%%%%%%%%%%%%%%%%%%%%

% Don't change these lines
\bsp	% typesetting comment
\label{lastpage}
\end{document}